\documentclass[preprint]{revtex4}
\usepackage{enumerate}
\usepackage{graphicx}
\usepackage{dcolumn}
\usepackage{bm}
\usepackage{amsmath}
\usepackage{amsxtra}
\usepackage{amstext}
\usepackage{amssymb}
\usepackage{latexsym}

\begin{document}

\title{
{ Nonlinear Closure Relations Theory for Transport Processes in Non-Equilibrium Systems}
}

\author{ Giorgio SONNINO}
\email{gsonnino@ulb.ac.be}
\affiliation{
Universit{\'e} Libre de Bruxelles (U.L.B.)\\
Department of Theoretical Physics and Mathematics\\
Campus de la Plaine C.P. 231 - Bvd du Triomphe\\
B-1050 Brussels - Belgium\\
}


\begin{abstract}

A decade ago, a macroscopic theory for closure relations has been proposed for systems out of Onsager's region. This theory is referred to as the {\it thermodynamic field theory} (TFT). The aim of the work was to determine the nonlinear flux-force relations that respect the thermodynamic theorems for systems far from equilibrium. We propose a new formulation of the TFT where one of the basic restrictions, namely the closed-form solution for the skew-symmetric piece of the transport coefficients, has been removed. In addition, the general covariance principle is replaced by the De Donder-Prigogine thermodynamic covariance principle (TCP). The introduction of TCP requires the application of an appropriate mathematical formalism, which is referred to as the {\it entropy-covariant formalism}. 
By geometrical arguments, we prove the validity of the Glansdorff-Prigogine Universal Criterion of Evolution. A new set of closure equations determining the nonlinear corrections to the linear ("Onsager") transport coefficients is also derived. The geometry of the thermodynamic space is non-Riemannian. However, it tends to be Riemannian for high values of the entropy production. In this limit, we recover the transport equations found by the old theory. Applications of our approach to transport in magnetically confined plasmas, materials submitted to temperature and electric potential gradients or to unimolecular triangular chemical reactions can be found at references cited herein. Transport processes in tokamak plasmas are of particular interest. In this case, even in absence of turbulence, the state of the plasma remains close to (but, it is not in) a state of local equilibrium. This prevents the transport relations from being linear.

\end{abstract}

\maketitle


\section{Introduction}

It is well known that the basic theory of dynamical systems should provide with an algorithm for the determination of the moments of the particle distribution functions $f^\alpha$ (i.e., the average values of the power of particle momenta ${\bf p}$), which are determined by the (fluctuating) fields through the kinetic equations. In the case of turbulent plasmas, for example, the most fundamental approach is the study of the stochastic kinetic equation coupled to the stochastic Maxwell equations. Such a self-consistent theory should not require any arbitrary assumption: it should produce equations of evolution for all the moments. In practice, however, an exact solution of this problem is impossible. Indeed, the equations of evolution of the moments have a hierarchical structure: the determination of a moment of order $n$ requires the knowledge of order $n+1$. Hence, the equations for the third moments will involve the fourth moments, and so on {\it ad infinitum}. Because of these difficulties, the fundamental studies, in spite of their basic importance, can not easily produce explicit results that can be directly compared to experiments. In order to obtain such results, one is led to make compromises: we must introduce additional simplifying assumptions allowing {\it to truncate} the hierarchy. As a result, we obtain a set of dynamical moments equations with a number of undetermined quantities: the equations are {\it not closed}. These quantities are of four kinds: thermodynamic quantities (such as temperature, pressure etc.), electromagnetic fields, moments-and energy-exchanges (such as the collisional friction forces or the collisional particles heat exchange) and fluxes (such as, the particle flux, the heat flux etc.). The dynamics of a thermodynamic system is finally based on the set of balance equations coupled to a (macroscopic) {\it theory for the closure relations}. Thus, in a macroscopic picture of thermodynamic systems, the formulation of a theory for the closure relations plays a fundamental role. The connection between the macroscopic equation and a microscopic distribution of particles should be established analyzing case by case (for example, for magnetically confined plasmas, see ref. \cite{sonnino3} and section \ref{isoef} - subsection {\it The Nonlinear Closure Equations}). 
 
\noindent The most important closure relations are the so-called {\it transport equations}, relating the dissipative fluxes to the thermodynamic forces that produce them. The study of these relations is the object of non-equilibrium thermodynamics. Close to equilibrium, the transport equations of a thermodynamic system are provided by the well-known Onsager theory. Indicating with $X^\mu$ and $J_\mu$ the thermodynamic forces and fluxes, respectively, the Onsager relations read
\begin{equation}\label{i1}
J_\mu=\tau_{0\mu\nu}X^\nu
\end{equation}
\noindent where $\tau_{0\mu\nu}$ are the transport coefficients. We suppose that all quantities involved in Eqs~(\ref{i1}) are written in dimensionless form. In this equation, as in the remainder of this paper, the Einstein summation convention on the repeated indexes is adopted. Matrix $\tau_{0\mu\nu}$ can be decomposed into a sum of two matrices, one symmetric and the other skew-symmetric, which we denote with $L_{\mu\nu}$ and $f_{0\mu\nu}$, respectively. The second principle of thermodynamics imposes that $L_{\mu\nu}$ be a positive definite matrix. The most important property of Eqs~(\ref{i1}) is that near equilibrium, the coefficients $\tau_{\mu\nu}$ are independent of the thermodynamic forces, so that
\begin{equation}\label{i2}
\frac{\partial\tau_{0\mu\nu}}{\partial X^\lambda}=0
\end{equation}
\noindent The region where Eqs~(\ref{i2}) hold, is called {\it Onsager's region} or, the {\it linear region}. A well-founded microscopic explanation on the validity of the linear phenomenological laws was developed by Onsager in 1931 \cite{onsager}. Onsager's theory is based on three assumptions: i) {\it The probability distribution function for the fluctuations of thermodynamic quantities} (Temperature, pressure, degree of advancement of a chemical reaction etc.) {\it is a Maxwellian} ii) {\it Fluctuations decay according to a linear law} and iii) {\it The principle of the detailed balance} (or the microscopic reversibility) {\it is satisfied}.  Onsager showed the equivalence of Eqs (\ref{i1}) and (\ref{i2}) with the assumptions i)-iii) [assumption iii) allows deriving the {\it reciprocity relations} $\tau_{0\mu\nu}=\tau_{0\nu\mu}$]. The Onsager theory of fluctuations starts from the Einstein formula linking the probability of a fluctuation, $\mathcal W$, with the entropy change, $\Delta S$, associated with the fluctuations from the state of equilibrium
\begin{equation}\label{i2a}
\mathcal{W}=W_0\exp[\Delta S/k_B]
\end{equation}
\noindent In Eq.~(\ref{i2a}), $k_B$ is the Bolzmann constant and $W_0$ is a normalization constant that ensures the sum of all probabilities equals one. The first assumption in the Onsager theory consists in postulating that the entropy variation is a bilinear expression of fluctuations. 

Many important theorems have been demonstrated for thermodynamic systems in the linear region. Among them, the most important one is the {\it Minimum Entropy Production Theorem}, showed by Prigogine in 1947 \cite{prigogine}. This theorem establishes that, in the Onsager region, for $a-a$ or $b-b$ processes (i.e., when the Onsager matrix is symmetric; see also the definition of $a-a$ and $b-b$ processes reported in the footnote \footnote{Here, we adopt the De Groot-Mazur terminology \cite{degroot}: the state of the system can be described by a number of independent variables. In general one distinguishes two types of macroscopic variables. The variables of the first type, denoted by the symbol $a$, are {\it even} functions of the particle velocities. The other variables, denoted by the symbol $b$, are {\it odd} functions of the particle velocities. Thermodynamic processes involving only variables $a$ ($b$) are referred to as {\it $a-a$ ($b-b$) processes}. It is possible to show that the Onsager reciprocal relations read $L^{a-a}_{\mu\nu}=L^{a-a}_{\nu\mu}$, $L^{b-b}_{\mu\nu}=L^{b-b}_{\nu\mu}$ and $L^{a-b}_{\mu\nu}=-L^{b-a}_{\nu\mu}$. \label{process}}), a thermodynamic system relaxes towards a steady-state in such a way that the rate of the entropy production is negative
\begin{equation}\label{i3}
\frac{d\sigma}{dt}\leq0\qquad \Bigl(\frac{d\sigma}{dt}=0\ \ {\rm at\ the\ steady\ state}\Bigr)
\end{equation}
\noindent where $\sigma=L_{\mu\nu}X^\mu X^\nu$ indicates the entropy production per unit volume and $t$ is time. Prigogine generalized Eq.~(\ref{i2a}), which applies only to adiabatic or isothermal transformations, by introducing the entropy production due to fluctuations. Denoting by $\xi_i$ ($i=1\cdots m$) the $m$ deviations of the thermodynamic quantities from their equilibrium value, Prigogine proposed that the probability distribution of finding a state in which the values $\xi_i$ lie between $\xi_i$ and $\xi_i+d\xi_i$ is given by \cite{prigogine}
\begin{equation}\label{i3a}
\mathcal{W}=W_0\exp[\Delta_{\rm I}  S/k_B]\qquad\quad
{\rm where}\qquad \Delta_{\rm I}   S=\int_E^F d_{\rm I} s\quad  {\rm ;}\quad \frac{d_{\rm I}  s}{dt}\equiv\int_\Omega\sigma dv
\end{equation}
\noindent $dv$ is a (spatial) volume element of the system, and the integration is over the entire space $\Omega$ occupied by the system in question. $E$ and $F$ indicate the equilibrium state and the state to which a fluctuation has driven the system, respectively. Note that this probability distribution remains unaltered for flux-force transformations leaving invariant the entropy production. 

\noindent In 1954, Glansdorff and Prigogine demonstrated a more general theorem, valid also when the system is out of Onsager's region \cite{prigogine1}. They showed that, regardless of the type of processes, a thermodynamic system relaxes towards a steady-state in such a way that the following quantity $\mathcal{P}$ is negative 
\begin{equation}\label{i4}
\mathcal{P}\equiv J_\mu\frac{d X^\mu}{dt}\leq0\qquad \Bigl(\mathcal{P}=0\ \ {\rm at\ the\ steady\ state}\Bigr)
\end{equation}
\noindent Inequality (\ref{i4}) reduces to inequality (\ref{i3}) for $a-a$ or $b-b$ processes in the Onsager region. For spatially-extended systems, the expression in Eqs.~(\ref{i4}) should be replaced by 
\begin{equation}\label{i5}
\mathcal{P}\equiv\int_\Omega {\mathcal J}_\mu\frac{d {\mathcal X}^\mu}{dt}dv\leq0\qquad \Bigl(\mathcal{P}=0\ \ {\rm at\ the\ steady\ state}\Bigr)
\end{equation}
\noindent ${\mathcal J}_\mu({\bf r},t)$ and ${\mathcal X}^\mu({\bf r},t)$ denote the space-time dependent fluxes and forces, respectively. The phenomenological equations are not needed for deriving this more general theorem and no restrictions are imposed to the transport coefficients (apart from the validity of the second principle of thermodynamics). Therefore, no use is made of the Onsager reciprocal relations, nor it is necessary to assume that the phenomenological coefficients are constants. The inequality expressed in (\ref{i4}) [or in (\ref{i5})] is referred to as the {\it Universal Criterion of Evolution} and it is the most general result obtained up to now in thermodynamics of irreversible processes. Out of Onsager's region, the transport coefficients may depend on the thermodynamic forces and Eqs~(\ref{i2}) may loose their validity. This happens when the first end/or the second assumption of the Onsager theory [i.e., the above-mentioned assumption 1) end/or assumption 2)]  are/is not satisfied. Magnetically confined tokamak plasmas are a typical example of thermodynamic systems out of Onsager's region. In this case, even in absence of turbulence, the local distribution functions of species (electrons and ions) deviate from the (local) Maxwellian. After a short transition time, the plasma remains close to (but, it is not in) a state of local equilibrium (see, for example, \cite{balescu2} and section \ref{isoef} - subsection {\it The Nonlinear Closure Equations}).

Transport in the nonlinear region, has been largely studied both experimentally and theoretically. In particular, many theories, based on the Fourier expansion of the transport coefficients in terms of the thermodynamic forces, have been proposed (see, for example, refs \cite{gyarmati}, \cite{li} and \cite{rysselberghe}). The theoretical predictions are however in disagreement with the experiments and this is mainly due to the fact that, in the series expansion, the terms of superior order are greater than those of inferior order. Therefore truncation of the series at some order is not mathematically justified. 

\noindent A thermodynamic field theory (TFT) has been developed in 1999 for proposing a closure relations theory for thermodynamic systems out of the Onsager region \cite{sonnino1}. In particular, the main objective of this work is to determine {\it how the linear flux-force relations} [i.e., Eqs~(\ref{i1})] {\it should be "deformed" in such a way that the thermodynamic theorems for systems far from equilibrium are respected} \cite{sonnino}. The Onsager coefficients enter in the theory as an input in the equations and they have to be calculated by kinetic theory. Attempts to derive a generally covariant thermodynamic field theory (GTFT) can be found in refs \cite{sonnino1}. The characteristic feature of the TFT is its purely macroscopic nature. This does not mean a formulation based on the macroscopic evolution equations, but rather a purely thermodynamic formulation starting solely from the entropy production and from the transport equations. The latter provide the possibility of defining an abstract space (the thermodynamic space), covered by the $n$ independent thermodynamic forces $X^\mu$, whose metric is identified with the symmetric part of the transport matrix. The law of evolution is not the dynamical law of particle motion, or the set of two-fluid macroscopic equations of plasma dynamics. The evolution in the thermodynamical forces space is rather determined by postulating three purely geometrical principles: {\it the shortest path principle}, {\it the skew-symmetric piece of the transport coefficients in closed form}, and {\it the principle of least action}. From theses principles, a set of closure equations, constraints, and boundary conditions are derived. These equations determine the nonlinear corrections to the linear ("Onsager") transport coefficients. However, the formulation of the thermodynamic field theory, as reported in refs~\cite{sonnino}, raises the following fundamental objection:
\vskip0.2truecm
\noindent {\it There are no strong experimental evidences supporting the requirement that the skew-symmetric piece of the transport coefficients is in a closed form}.
\vskip0.2truecm
\noindent Moreover, the principle of general covariance, which in refs \cite{sonnino1} has been  assumed to be valid for general transformations in the space of thermodynamic configurations, is, in reality, respected only by a very limited class of thermodynamic processes. In this paper, through an appropriate mathematical formalism, the {\it entropy-covariant formalism}, the entire TFT is re-formulated removing the assumptions regarding the closed-form of the skew-symmetric piece of the transport coefficients and the general covariance principle (GCP). The GCP is replaced by  the {\it thermodynamic covariance principle}  (TCP), or the De Donder-Prigogine statement \cite{dedonder}-\cite{prigogine2}, establishing that thermodynamic systems, obtained by a transformation of forces and fluxes in such a way that the entropy production remains unaltered, are thermodynamically equivalent. This principle applies to transformations in the thermodynamic space and they may be referred to as the {\it  thermodynamic coordinate transformations} (TCT). It is worthwhile mentioning that the TCP is actually largely used in a wide variety of thermodynamic processes ranging from non equilibrium chemical reactions to transport processes in tokamak plasmas (see, for examples, the papers and books cited in refs \cite{balescu2} and \cite{hinton}). To the author knowledge, the validity of the thermodynamic covariance principle has been verified empirically without exception in physics until now.  

\noindent The analysis starts from the following observation. Consider a relaxation process of a thermodynamic system in the Onsager region. If the system relaxes towards a steady-state along the shortest path in the thermodynamic space, then the Universal Criterion of Evolution is automatically satisfied. Indeed, in this case, we can write 
\begin{equation}\label{i6}
J_\mu{\dot X}^\mu=(L_{\mu\nu}+f_{0\mu\nu})X^\nu{\dot X^\mu}
\end{equation}
\noindent where the dot over the variables indicates the derivative with respect to the arc parameter $\varsigma$, defined as 
\begin{equation}\label{i7}
d\varsigma^2=L_{\mu\nu}dX^\mu dX^\nu
\end{equation}
\noindent Parameter $\varsigma$ can be chosen in such a way that it vanishes when the system begins to evolve and it assumes the value, say $l$, when the system reaches the steady-state. In the Onsager region, the thermodynamic space is an Euclidean space with metric $L_{\mu\nu}$. The equation of the shortest path reads ${\ddot X}^\mu=0$, with solution of the form
\begin{equation}\label{i8}
X^\mu=a^\mu\varsigma +b^\mu
\end{equation}
\noindent where $a^\mu$ and $b^\mu$ are arbitrary constant independent of the arc parameter. Inserting Eq.~(\ref{i8}) into Eq.~(\ref{i6}) and observing that  $L_{\mu\nu}a^\mu a^\nu=1$ and $f_{0\mu\nu}a^\mu a^\nu=0$, we find
\begin{equation}\label{i9}
J_\mu{\dot X}^\mu=\varsigma+\tau_{0\mu\nu}a^\mu b^\nu
\end{equation}
\noindent At the steady state (i.e. for $\varsigma=l$) $J_\mu{\dot X}^\mu\mid_{st.state}=0$. Eq.~(\ref{i9}) can then be written as
\begin{equation}\label{i10}
P=-(l-\varsigma)\leq0\qquad({\rm with}\quad P\equiv J_{\mu}{\dot X}^\mu)
\end{equation}
\noindent or
\begin{equation}\label{i11}
{\mathcal P}=-(l-\varsigma)\Bigl(L_{\mu\nu}\frac{dX^\mu}{dt}\frac{dX^\nu}{dt}\Bigr)^{1/2}\leq0
\end{equation}
\noindent The equation for the dissipative quantity $P$, when the thermodynamic system relaxes in the linear region, is thus given by Eq.~(\ref{i9}):
\begin{equation}\label{i12}
\frac{dP}{d\varsigma}=1
\end{equation}
\noindent Also note that ${\dot\sigma}=2P\leq0$ i.e., the minimum entropy production theorem is also satisfied during relaxation. Now, our question is: "{\it How can we "deform" the linear flux-force relations in such a way that the Universal Criterion of Evolution remains automatically satisfied, without imposing any restrictions to the transport coefficients, also out of Onsager's region ?}". Outside the linear region, one may be tempted to construct a Riemannian space (of $3$ or more dimensions) which is projectively flat i.e., having a vanishing Weyl's projective curvature tensor. In this case, indeed, there exists a coordinate system such that the equations of the shortest path are linear in the coordinates [i.e., the shortest paths are given by equations of the form (\ref{i8})]. In this respect, we have the following Weyl theorem \cite{weyl}: a necessary and sufficient condition that a Riemannian space be projectively flat is that its Riemannian curvature be constant everywhere. On the other hand, to re-obtain the Onsager relations, we should also require that, near equilibrium, the Riemannian space reduces to a flat space (which has zero Riemannian curvature). The Weyl theorem can be conciliated with our request only if there exists a coordinate system such that Eqs~(\ref{i2}) are valid everywhere, which is in contrast with experiments. Thus one wants the Universal Criterion of Evolution satisfied also out of the Onsager region, without imposing a priori {\it any} restrictions on transport coefficients, a {\it non-Riemannian} thermodynamic space is required. Clearly, a transport theory without a knowledge of microscopic dynamical laws can not be developed. Transport theory is only but an aspect of non-equilibrium statistical mechanics, which provides the link between micro and macro-levels. This link appears indirectly in the "unperturbed" matrices, i.e. the $L_{\mu\nu}$ and the $f_{0\mu\nu}$ coefficients, used as an input in the equations. These coefficients, which depend on the specific material under consideration, have to be calculated in the usual way by kinetic theory. 

\noindent In section \ref{isoef}, we introduce a non-Riemannian space whose geometry is constructed in such a way that
\begin{description}
\item {\it {\bf A.} The theorems, valid when a generic thermodynamic system relaxes out of equilibrium, are satisfied};
\item {\it {\bf B.} The nonlinear closure equations are covariant under the thermodynamic coordinate transformations (TCT)}.
\end{description}
\noindent We shall see that the properties of geometry do not depend on the shortest paths but upon {\it a particular expression of the affine connection}. Our geometry is then of {\it affine type} and not of projective type. At the end of section \ref{isoef}, we derive the nonlinear closure equations through an appropriate mathematical formalism: the {\it entropy (production)-covariant formalism} (in the sequel, the {\it entropy-covariant formalism}). This formalism allows to respect the De Donder-Prigogine statement. New geometrical objects like {\it thermodynamic covariant differentiation} or the {\it thermodynamic curvature} are also introduced. We shall see that under the weak-field approximation and when $\sigma\gg 1$, but only in these limits, the new nonlinear closure equations reduce to the ones obtained in refs \cite{sonnino}. So that, all results found in refs \cite{sonnino3}, for magnetically confined plasmas, and in refs \cite{sonnino4}, for the nonlinear thermoelectric effect and the unimolecular triangular reaction, remain valid. In section \ref{ttse} we show that this formalism is able to verify the thermodynamic theorems (in particular, the Universal Criterion of Evolution) for systems relaxing out of the Onsager region. Mathematical details and demonstrations of the theorems are reported in the annexes.

\noindent It should be noted that, geometrical formalisms have been applied for treating topics different to the transport closure theory, such as the use of the matrix of the second derivatives of the entropy as a metric tensor in the analysis of fluctuations (see, for example, \cite{ruppeiner}) and the use of symplectic geometries in the analysis of nonlinear evolution equations of dynamical systems \cite{grmela}. 

\vskip 0.2truecm
\section{The Entropy-Covariant Formalism}\label{isoef}
\vskip 0.2truecm
\noindent Consider a thermodynamic system driven out from equilibrium by a set of $n$ independent thermodynamic forces $\{X^\mu\}$ ($\mu=1,\cdots n$). It is also assumed that the system is submitted to time-independent boundary conditions. The set of conjugate flows, $\{J_{\mu}\}$, is coupled to the thermodynamic forces through the relation 
\begin{equation}\label{ief0}
J_{\mu}=\tau_{\mu\nu}(X)X^\nu
\end{equation}
\noindent where the transport coefficients, $\tau_{\mu\nu}(X)$, may depend on the thermodynamic forces. The symmetric piece of $\tau_{\mu\nu}(X)$ is denoted with $g_{\mu\nu}(X)$ and the skew-symmetric piece as $f_{\mu\nu}(X)$:
\begin{equation}\label{ief1}
\tau_{\mu\nu}(X)=\frac{1}{2}[\tau_{\mu\nu}(X)+\tau_{\nu\mu}(X)]+\frac{1}{2}[\tau_{\mu\nu}(X)-\tau_{\nu\mu}(X)]=g_{\mu\nu}(X)+f_{\mu\nu}(X)
\end{equation}
\noindent where
\begin{eqnarray}\label{ief2}
&&g_{\mu\nu}(X)=\frac{1}{2}[\tau_{\mu\nu}(X)+\tau_{\nu\mu}(X)]=g_{\nu\mu}(X)\\
&&f_{\mu\nu}(X)=\frac{1}{2}[\tau_{\mu\nu}(X)-\tau_{\nu\mu}(X)]=-f_{\nu\mu}(X)
\end{eqnarray}
\noindent  It is assumed that $g_{\mu\nu}(X)$ is a positive definite matrix. For conciseness, in the sequel we drop the symbol $(X)$ in $\tau_{\mu\nu}$, $g_{\mu\nu}$ and $f_{\mu\nu}$, being implicitly understood that these matrices may depend on the thermodynamic forces. With the elements of the transport coefficients two objects are constructed: {\it operators}, which may act on thermodynamic tensorial objects and {\it thermodynamic tensorial objects}, which under coordinate (forces) transformations, obey to well specified transformation rules.
\vskip 0.2truecm
\noindent{\bf Operators}
\vskip 0.2truecm
Two operators are introduced, the {\it entropy production operator} $\sigma (X)$ and the {\it dissipative quantity operator} ${\tilde P}(X)$, acting on the thermodynamic forces in the following manner
\begin{eqnarray}\label{o1}
&&\sigma (X):\rightarrow\sigma(X)\equiv X gX^T\nonumber\\
&& {\tilde P}(X):\rightarrow {\tilde P}(X)\equiv X\tau \Bigl[\frac{dX}{d\varrho}\Bigr]^T
\end{eqnarray}
\noindent In Eqs~(\ref{o1}), the transport coefficients are then considered as elements of the two $n$ x $n$ matrices, $\tau$ and $g$. The positive definiteness of the matrix $g_{\mu\nu}$ ensures the validity of the second principle of thermodynamics: $\sigma\ge0$. These matrices multiply the thermodynamic forces $X$ expressed as $n$ x $1$ column matrices. We already anticipate that parameter $\varrho$, defined in Eq.~(\ref{tt2b}), is invariant under the thermodynamic coordinate transformations. Thermodynamic states $X_{s}$ such that
\begin{eqnarray}\label{o2}
\Big[{\tilde P}(X)\frac{d\varrho}{dt}\Big]_{X=X_{s}}
\!\!\!\!=0
\end{eqnarray}
\noindent are referred to as {\it steady-states}. Of course, the steady-states should be invariant expressions under the thermodynamic coordinate transformations. Eqs~(\ref{o1}) {\it should not} be interpreted as the metric tensor $g_{\mu\nu}$, which acts on the coordinates. The metric tensor {\it acts only on} elements of the tangent space (like $dX^\mu$, see the forthcoming paragraphs) or on the thermodynamic tensorial objects.
\vskip 0.2truecm
\noindent{\bf Transformation Rules of Entropy Production, Forces, and Flows}
\vskip 0.2truecm
According to the De Donder-Prigogine statement \cite{dedonder}, \cite{prigogine2} {\it thermodynamic systems are thermodynamically equivalent if, under transformation of fluxes and forces the bilinear form of the entropy production, $\sigma$, remains unaltered} \footnote{In some examples of chemical reactions, the only condition of invariance of entropy production may not be sufficient to assure the equivalent character of two descriptions ($J_\mu, X^\mu$) and ($J'_\mu, X'^\mu$).  In ref.\cite{prigogine} we can find the case where it is also necessary to impose additional invariances of the rate of change of the number of moles. This is necessary to avoid certain paradoxes to which Verschaffelt \cite{verschaffelt} has called attention (cf., also \cite{davies}). \label{process}}. In mathematical terms, this implies:
 \begin{equation}\label{tr1}
 \sigma=J_\mu X^\mu=J'_\mu X'^\mu
 \end{equation}
 \noindent This condition and the condition that also the dissipative quantity [cf. Eqs~(\ref{o1})] must be an invariant expression require that the transformed thermodynamic forces and flows satisfy the relation  
\begin{eqnarray}\label{tr2}
&&X'^\mu=\frac{\partial X'^{\mu}}{\partial X^\nu} X^\nu\nonumber\\
&& J'_\mu=\frac{\partial X^{\nu}}{\partial X'^\mu}J_\nu
\end{eqnarray}
\noindent These transformations may be referred to as {\it Thermodynamic Coordinate Transformations} (TCT). The expression of entropy production becomes accordingly
 \begin{equation}\label{tr3}
 \sigma=J_\mu X^\mu=\tau_{\mu\nu}X^\mu X^\nu=g_{\mu\nu}X^\mu X^\nu=g'_{\mu\nu}X'^\mu X'^\nu=\sigma'
 \end{equation}
 \noindent From Eqs~(\ref{tr2}) and (\ref{tr3}) we derive
 \begin{equation}\label{tr4}
g'_{\lambda\kappa}=g_{\mu\nu}\frac{\partial X^\mu}{\partial X'^\lambda}
 \frac{\partial X^\nu}{\partial X'^\kappa}
\end{equation}
\noindent Moreover, inserting Eqs~(\ref{tr2}) and Eq.~(\ref{tr4}) into relation $J_{\mu}=(g_{\mu\nu}+f_{\mu\nu})X^\nu$, we obtain
\begin{equation}\label{tr5}
J'_\lambda=\Bigl(g'_{\lambda\kappa}+f_{\mu\nu}\frac{\partial X^\mu}{\partial X'^\lambda}\frac{\partial X^\nu}{\partial X'^\kappa}\Bigr)X'^\kappa
\end{equation}
\noindent or 
 \begin{equation}\label{tr6}
J'_\lambda=(g'_{\lambda\kappa}+f'_{\lambda\kappa})X'^\kappa\quad\qquad{\rm with}\qquad
f'_{\lambda\kappa}=f_{\mu\nu}\frac{\partial X^\mu}{\partial X'^\lambda}
 \frac{\partial X^\nu}{\partial X'^\kappa}
\end{equation}
\noindent Hence, the transport coefficients transform like a {\it thermodynamic tensor of second order}
\footnote{We may qualify as {\it thermodynamic tensor} or, simply {\it thermo-tensor}, (taken as a single noun) a set of quantities where {\it only transformations Eqs~(\ref{tr2}) are involved}. This is in order to qualify as a {\it tensor}, a set of quantities, which satisfies certain laws of transformation when the coordinates undergo a general transformation. Consequently every tensor is a thermodynamic tensor but the converse is not true.\label{thermotensor}}.

\vskip0.5truecm
\noindent{\bf Properties of the TCT}
\vskip0.5truecm

\noindent By direct inspection, it is easy to verify that the general solutions of equations (\ref{tr2}) are 
\begin{equation}\label{tct1}
X'^\mu=X^1F^\mu\Bigl(\frac{X^2}{X^1},\ \frac{X^3}{X^2},\ \cdots\ \frac{X^n}{X^{n-1}}\Bigr)
\end{equation}
\noindent where $F^\mu$ are {\it arbitrary functions} of variables $X^j/X^{j-1}$ with ($j=2,\dots, n$). Hence, the TCT may be {\it highly nonlinear coordinate transformations} but, in the Onsager region, we may (or we must) require that they have to reduce to 
\begin{equation}\label{tct1a}
X'^\mu=c_\nu^\mu X^\nu
\end{equation}
\noindent where $c_\nu^\mu$ are constant coefficients (i.e., independent of the thermodynamic forces). We note that from Eq.(\ref{tr2}), the following important identities are derived
\begin{equation}\label{tct2}
X^\nu\frac{\partial^2X'^\mu}{\partial X^\nu\partial X^\kappa}=0\qquad ;\qquad X'^\nu\frac{\partial^2X^\mu}{\partial X'^\nu\partial X'^\kappa}=0
\end{equation}
\noindent Moreover
\begin{eqnarray}\label{tct3}
&&dX'^\mu=\frac{\partial X'^{\mu}}{\partial X^\nu} dX^\nu \nonumber\\
&&\frac{\partial}{\partial X'^\mu}= \frac{\partial X^{\nu}}{\partial X'^\mu}\frac{\partial}{\partial X^\nu} 
\end{eqnarray}
\noindent i.e., $dX^\mu$ and $\partial/\partial X^{\mu}$ transform like a thermodynamic contra-variant and a thermodynamic covariant vector, respectively. According to Eq.~(\ref{tct3}), thermodynamic vectors $dX^\mu$ define the {\it tangent space} to $Ts$. It also follows that the operator $P(X)$, i.e. the dissipation quantity, and in particular the definition of steady-states, are invariant under TCT. Parameter $\varsigma$, defined as 
\begin{equation}\label{tct4}
d\varsigma^2=g_{\mu\nu}dX^\mu dX^\nu
\end{equation}
\noindent is a scalar under TCT. The operator $\mathcal{O}$
\begin{equation}\label{pts14d}
\mathcal{O}\equiv X^\mu\frac{\partial}{\partial X^\mu}=X'^\mu\frac{\partial}{\partial X'^\mu}=\mathcal{O}'
\end{equation}
\noindent is also invariant under TCT. This operator plays an important role in the formalism. 
\vskip 0.2truecm
\noindent{\bf Thermodynamic Space, Thermodynamic Covariant Derivatives and Thermodynamic Curvature}
\vskip 0.2truecm
A non-Riemannian space with an affine connection $\Gamma_{\alpha\beta}^\mu$ is now introduced (see also Appendix \ref{dictionary}). Consider an $n$-space in which the set of quantities $\Gamma_{\alpha\beta}^\mu$ is assigned as functions of the $n$ independent thermodynamic forces $X^\mu$, chosen as coordinate system. Under a coordinate (forces) transformation, it is required that the functions $\Gamma_{\alpha\beta}^\mu$ transform according to the law
\begin{equation}\label{ts1}
{\Gamma'}_{\alpha\beta}^\mu=\Gamma_{\lambda\kappa}^\nu\frac{\partial X'^\mu}{\partial X^\nu}\frac{\partial X^\lambda}{\partial X'^\alpha}\frac{\partial X^\kappa}{\partial X'^\beta}+\frac{\partial X'^\mu}{\partial X^\nu}\frac{\partial^2X^\nu}{\partial X'^\alpha \partial X'^\beta}
\end{equation}
\noindent With the linear connection $\Gamma_{\alpha\beta}^\mu$, the absolute derivative of an arbitrary thermodynamic contra-variant vector, denoted by $T^\mu$, along a curve can be defined as
\begin{equation}\label{ts2}
\frac{\delta T^\mu}{\delta\varsigma}=\frac{dT^\mu}{d\varsigma}+\Gamma^\mu_{\alpha\beta}T^\alpha\frac{dX^\beta}{d\varsigma}
\end{equation}
\noindent It is easily checked that, if the parameter along the curve is changed from $\varsigma$ to $\varrho$, then the absolute derivative of a thermodynamic tensor field with respect to $\varrho$ is $d\varsigma/d\varrho$ times the absolute derivative with respect to $\varsigma$. The absolute derivative of any contra-variant thermodynamic tensor may be easily obtained generalizing Eq.~(\ref{ts2}). In addition, the linear connection $\Gamma^\mu_{\alpha\beta}$ is submitted to the following basic postulates:
\begin{description}
\item{\it {\bf 1.} The absolute derivative of a thermodynamic contra-variant tensor is a thermodynamic tensor of the same order and type.}
\item{\it {\bf 2.} The absolute derivative of an outer product of thermodynamic tensors, is given, in terms of factors, by the usual rule for differentiating a product.}
\item{\it {\bf 3.} The absolute derivative of the sum of thermodynamic tensors of the same type is equal to the sum of the absolute derivatives of the thermodynamic tensors.}
\end{description}
\noindent In a space with a linear connection, we can introduce the notion of the {\it shortest path} defined as {\it a curve such that a thermodynamic vector, initially tangent to the curve and propagated parallelly along it, remains tangent to the curve at all points}. By a suitable choice of the parameter $\varrho$, the differential equation for the shortest path is simplified reducing to
\begin{equation}\label{ts3}
\frac{d^2X^\mu}{d\varrho^2}+\Gamma^\mu_{\alpha\beta}\frac{dX^\alpha}{d\varrho}\frac{dX^\beta}{d\varrho}=0
\end{equation}
\noindent To satisfy the general requirement ${\bf A.}$, (see section \ref{ttse}), it is required that the absolute derivative of the entropy production satisfies the equality 
\begin{equation}\label{ts4}
\frac{\delta\sigma}{\delta\varsigma}=J_\mu\frac{\delta X^\mu}{\delta\varsigma}+X^\mu\frac{\delta J_\mu}{\delta\varsigma}
\end{equation}
\noindent More in general, it is required that the operations of contraction and absolute differentiation commute for all thermodynamic vectors. As a consequence, the considered space should be a space with a single connection. The absolute derivative of an arbitrary covariant thermodynamic vector, denoted by $T_\mu$, is then defined as 
\begin{equation}\label{ts5}
\frac{\delta T_\mu}{\delta\varsigma}=\frac{dT_\mu}{d\varsigma}-\Gamma^\alpha_{\mu\beta}T_\alpha\frac{dX^\beta}{d\varsigma}
\end{equation}
\noindent The absolute derivative of the most general contra-variant, covariant and mixed thermodynamic tensors may be obtained generalizing Eqs~(\ref{ts2}) and (\ref{ts5}). The derivatives, covariant under TCT, of thermodynamic vectors, are defined as
\begin{eqnarray}\label{ts6}
&&T^\mu_{\mid\nu}=\frac{\partial T^\mu}{\partial X^\nu}+\Gamma^\mu_{\alpha\nu}T^\alpha\nonumber\\
&&T_{\mu\mid\nu}=\frac{\partial T_\mu}{\partial X^\nu}-\Gamma^\alpha_{\mu\nu}T_\alpha
\end{eqnarray}
\noindent For the entropy production, it is also required that
\begin{equation}\label{ts7}
\sigma_{\mid\mu\mid\nu}=\sigma_{\mid\nu\mid\mu}
\end{equation}
\noindent More in general, Eq.~(\ref{ts7}) should be verified for any thermodynamic scalar $T$. This postulate requires that the linear single connection $\Gamma^\mu_{\alpha\beta}$ is also symmetric i.e., $\Gamma^\mu_{\alpha\beta}=\Gamma^\mu_{\beta\alpha}$. A non-Riemannian geometry can now be constructed out of $n^2(n+1)/2$ quantities, the components of $\Gamma^\mu_{\alpha\beta}$, according to the general requirements  ${\bf A}$ and ${\bf B}$ mentioned in the introduction.

\noindent In the forthcoming paragraph, the expression of the affine connection $\Gamma^\mu_{\alpha\beta}$ is determined from assumption ${\bf A}$. In section  \ref{ttse} it is shown that the Universal Criterion of Evolution, applied to thermodynamic systems relaxing towards a steady-state, is automatically satisfied along the shortest path if, in case of symmetric processes (i.e., for $a-a$ or $b-b$ processes), we impose
\begin{eqnarray}\label{ts8}
&&\!\!\!\!\!\!\!\!\!\!\!\!\!\!\!\!\!\!\!\!\!\!\!\!
\Gamma^\mu_{\alpha\beta}=\frac{1}{2}g^{\mu\lambda}\Bigl(\frac{\partial g_{\lambda\alpha}}{\partial X^\beta}+\frac{\partial g_{\lambda\beta}}{\partial X^\alpha}-\frac{\partial g_{\alpha\beta}}{\partial X^\lambda}\Bigr)+\frac{1}{2\sigma}X^\mu\mathcal{O}(g_{\alpha\beta})\\
&&\!\!\!\!\!\!\!\!\!\!\!\!\!\!\!\!\!\!\!\!\!\!\!\!
{\rm where}\qquad \mathcal{O}(g_{\alpha\beta})\equiv X^\eta\frac{\partial g_{\alpha\beta}}{\partial X^\eta}\nonumber
\end{eqnarray}
\noindent In the general case, we have 
\begin{eqnarray}\label{ts9}
\!\!\!\!\!\!\!\!\!\!\!\!\!\!\!\Gamma^\mu_{\alpha\beta}=\!\!\!\!&&{\check N}^{\mu\kappa}g_{\kappa\lambda}\begin{Bmatrix} 
\lambda \\ \alpha\beta
\end{Bmatrix}+\frac{{\check N}^{\mu\kappa}}{2\sigma}X_\kappa\mathcal{O}(g_{\alpha\beta})+\frac{{\check N}^{\mu\kappa} }{2\sigma}X_\kappa X^\lambda\Bigl(\frac{\partial f_{\alpha\lambda}}{\partial X^\beta}+\frac{\partial f_{\beta\lambda}}{\partial X^\alpha}\Bigr)\nonumber\\
&&
\qquad\qquad\qquad\ \ \!
+\frac{{\check N}^{\mu\kappa}}{2\sigma}f_{\kappa\varsigma}X^\varsigma X^\lambda\Bigl(\frac{\partial g_{\alpha\lambda}}{\partial X^\beta}+\frac{\partial g_{\beta\lambda}}{\partial X^\alpha}\Bigr)
\end{eqnarray}
\noindent where the thermodynamic Christoffel symbols of the second kind are introduced
\begin{equation}\label{ts10}
\begin{Bmatrix} 
\mu \\ \alpha\beta
\end{Bmatrix}=\frac{1}{2}g^{\mu\lambda}\Bigl(\frac{\partial g_{\lambda\alpha}}{\partial X^\beta}+\frac{\partial g_{\lambda\beta}}{\partial X^\alpha}-\frac{\partial g_{\alpha\beta}}{\partial X^\lambda}\Bigr)
\end{equation}
\noindent and matrix ${\check N}^{\mu\kappa}$ is defined as
\begin{equation}\label{ts11}
\!\ \!N_{\mu\nu}\equiv g_{\mu\nu}+\frac{1}{\sigma}f_{\mu\kappa}X^\kappa X_\nu+\frac{1}{\sigma}f_{\nu\kappa}X^\kappa X_\mu\quad{\rm with}\ \ \ {\check N^{\mu\kappa}}:\ \ {\check N}^{\mu\kappa}N_{\nu\kappa}=\delta_{\nu}^{\mu}
\end{equation}
\noindent In appendix \ref{appx} it is proven that the affine connections Eqs~(\ref{ts8}) and (\ref{ts9}) transform, under a TCT, as in Eq.~(\ref{ts1}) and satisfy the postulates ${\bf 1.}$, ${\bf 2.}$ and ${\bf 3.}$ From Eq.~(\ref{ts11}) we easily check that 
\begin{eqnarray}\label{ts12}
\!\!\!\!\!\!\!\!\!\!\!\!\!\!\!\!\!\!\!\!\!\!\!\!\!\!\!\!&&N_{\mu\nu}=N_{\nu\mu}\nonumber\\
\!\!\!\!\!\!\!\!\!\!\!\!\!\!\!\!\!\!\!\!\!\!\!\!\!\!\!\!&&N_{\mu\nu}X^\nu=\Bigl(g_{\mu\nu}+\frac{1}{\sigma}f_{\mu\kappa}X^\kappa X_\nu+\frac{1}{\sigma}f_{\nu\kappa}X^\kappa X_\mu \Bigr)X^\nu
=
J_{\mu}\nonumber\\
\!\!\!\!\!\!\!\!\!\!\!\!\!\!\!\!\!\!\!\!\!\!\!\!\!\!\!\!&&N_{\mu\nu}X^\mu=N_{\nu\mu}X^\mu=J_\nu\\
\!\!\!\!\!\!\!\!\!\!\!\!\!\!\!\!\!\!\!\!\!\!\!\!\!\!\!\!&&N_{\mu\nu} X^\nu X^\mu
=J_\mu X^\mu=\sigma\nonumber
\end{eqnarray}
\noindent While
\begin{eqnarray}\label{ts13}
\!\!\!\!\!\!\!\!\!\!\!\!\!\!\!\!\!\!\!\!\!\!\!\!\!\!\!\!\!\!\!\!\!\!\!\!\!\!\!\!\!\!\!\!\!\!\!\!\!\!\!\!\!\!\!\!\!\!\!
&&{\check N}^{\mu\kappa}={\check N}^{\kappa\mu}\nonumber\\
\!\!\!\!\!\!\!\!\!\!\!\!\!\!\!\!\!\!\!\!\!\!\!\!\!\!\!\!\!\!\!\!\!\!\!\!\!\!\!\!\!\!\!\!\!\!\!\!\!\!\!\!\!\!\!\!\!\!\!
&&{\check N}^{\mu\kappa}J_{\mu}={\check N}^{\mu\kappa}N_{\mu\nu}X^\nu={\check N}^{\kappa\mu}N_{\nu\mu}X^\nu=X^\kappa\\
\!\!\!\!\!\!\!\!\!\!\!\!\!\!\!\!\!\!\!\!\!\!\!\!\!\!\!\!\!\!\!\!\!\!\!\!\!\!\!\!\!\!\!\!\!\!\!\!\!\!\!\!\!\!\!\!\!\!\!
&&{\check N}^{\mu\kappa}J_{\kappa}={\check N}^{\kappa\mu}J_{\kappa}=X^\mu\nonumber\\
\!\!\!\!\!\!\!\!\!\!\!\!\!\!\!\!\!\!\!\!\!\!\!\!\!\!\!\!\!\!\!\!\!\!\!\!\!\!\!\!\!\!\!\!\!\!\!\!\!\!\!\!\!\!\!\!\!\!\!
&&{\check N}^{\mu\kappa}J_{\kappa}J_{\mu}=X^{\mu}J_{\mu}=\sigma\nonumber
\end{eqnarray}
\noindent At this point, we are confronted with the following theorem \cite{eisenhart} "{\it For two symmetric connections, the most general change which preserves the paths} {\it is}
\begin{equation}\label{ts14}
{\bar\Gamma}_{\alpha\beta}^\mu=\Gamma_{\alpha\beta}^\mu+\delta^\mu_\alpha\psi_\beta+\delta^\mu_\beta\psi_\alpha
\end{equation}
\noindent {\it where} $\psi_\alpha$ {\it is an arbitrary covariant thermodynamic vector and} $\delta^\mu_\nu$ {\it denotes the Kronecker tensor"}. In literature, modifications of the connection similar to Eqs~(\ref{ts14}) are referred to as {\it projective transformations} of the connection and $\psi_\alpha$ the {\it projective covariant vector}. The introduction of the affine connection gives rise, then, to the following difficulty:  the Universal Criterion of Evolution is satisfied for every shortest path constructed with affine connections ${\bar\Gamma}_{\alpha\beta}^\mu$, linked to $\Gamma_{\alpha\beta}^\mu$ by projective transformations. This leads to an indetermination of the expression for the affine connection, which is not possible to remove by using the De Donder-Prigogine statement and the thermodynamic theorems alone. This problem can be solved by postulating that {\it the nonlinear closure equations} (i.e., the equations for the affine connection and the transport coefficients) {\it be symmetric and projective-invariant} (i.e., invariant under projective transformations). 

\noindent For any arbitrary covariant thermodynamic vector field, denoted by  $T_\mu(X)$, we can form the thermodynamic tensor $R^{\mu}_{\nu\lambda\kappa}(X)$ in the following manner \cite{synge}
\begin{equation}\label{ts15}
T_{\nu\mid\lambda\mid\kappa}(X)-T_{\nu\mid\kappa\mid\lambda}(X)=T_{\mu}(X)R^{\mu}_{\nu\lambda\kappa}(X)
\end{equation}
\vskip 0.2truecm
\noindent where [by omitting, for conciseness, the symbol $(X)$] 
\begin{equation}\label{ts16}
R^{\mu}_{\nu\lambda\kappa}=\frac{\partial \Gamma^\mu_{\nu\kappa}}{\partial X^\lambda}-\frac{\partial \Gamma^\mu_{\nu\lambda}}{\partial X^\kappa}+\Gamma^\eta_{\nu\kappa}\Gamma^\mu_{\eta\lambda}-\Gamma^\eta_{\nu\lambda}\Gamma^\mu_{\eta\kappa}
\end{equation}
\noindent  with $R^{\mu}_{\nu\lambda\kappa}$ satisfying the following identities
\begin{eqnarray}\label{ts17}
&&R^{\mu}_{\nu\lambda\kappa}=-R^{\mu}_{\nu\kappa\lambda}\nonumber\\
&&R^{\mu}_{\nu\lambda\kappa}+R^{\mu}_{\lambda\kappa\nu}+R^{\mu}_{\lambda\nu\kappa}=0\\
&&R^{\mu}_{\nu\lambda\kappa\mid\eta}+R^{\mu}_{\nu\kappa\eta\mid\lambda}+R^{\mu}_{\nu\eta\lambda\mid\kappa}=0\nonumber
\end{eqnarray}
\noindent By contraction, we obtain two distinct thermodynamic tensors of second order
\begin{eqnarray}\label{ts18}
&&R_{\nu\lambda}=R^{\mu}_{\nu\lambda\mu}=\frac{\partial \Gamma^\mu_{\nu\mu}}{\partial X^\lambda}-\frac{\partial \Gamma^\mu_{\nu\lambda}}{\partial X^\mu}+\Gamma^\eta_{\nu\mu}\Gamma^\mu_{\eta\lambda}-\Gamma^\eta_{\nu\lambda}\Gamma^\mu_{\eta\mu}\nonumber\\
&&F_{\lambda\nu}=\frac{1}{2}R^{\mu}_{\mu\lambda\nu}=\frac{1}{2}\Bigl(\frac{\partial \Gamma^\mu_{\nu\mu}}{\partial X^\lambda}-\frac{\partial \Gamma^\mu_{\lambda\mu}}{\partial X^\nu}\Bigr)
\end{eqnarray} with $F_{\lambda\nu}$ being skew-symmetric and $R_{\nu\lambda}$ asymmetric. Tensor $R_{\nu\lambda}$ can be re-written as 
\begin{eqnarray}\label{ts19}
&&\!\!\!\!\!\!\!\!\!\!\!\!\!\!\!\!R_{\nu\lambda}=B_{\nu\lambda}+F_{\lambda\nu}\qquad{\rm where}\nonumber\\
&&\!\!\!\!\!\!\!\!\!\!\!\!\!\!\!\!B_{\nu\lambda}=B_{\lambda\nu}=\frac{1}{2}\Bigl(\frac{\partial \Gamma^\mu_{\nu\mu}}{\partial X^\lambda}+
\frac{\partial \Gamma^\mu_{\lambda\mu}}{\partial X^\nu}\Bigr)
-\frac{\partial \Gamma^\mu_{\nu\lambda}}{\partial X^\mu}+\Gamma^\eta_{\nu\mu}\Gamma^\mu_{\eta\lambda}-\Gamma^\eta_{\nu\lambda}\Gamma^\mu_{\eta\mu}
\end{eqnarray}
\noindent Hence, $F_{\lambda\nu}$ is the skew-symmetric part of $R_{\nu\lambda}$ \footnote{Of course, $R^{\mu}_{\nu\lambda\kappa}$ and $R_{\nu\lambda}$ do not coincide with the Riemannian curvature tensor and the Ricci tensor, respectively (see also Appendix \ref{comparison}).\label{tensors}}. It is argued that the closure equations can be derived by variation of a stationary action, which involves $R_{\nu\lambda}$. Symmetric and projective-invariant closure equations may be obtained by adopting the following strategy: 1) a suitable projective transformation of the affine connection is derived so that $R_{\nu\lambda}$ be symmetric and $F_{\lambda\nu}$ be a zero thermodynamic tensor and 2) the most general projective transformation that leaves unaltered $R_{\nu\lambda}$ and $F_{\lambda\nu}$ ($=0$) is determined. By a projective transformation, it is found that
\begin{eqnarray}\label{ts20}
&&{\bar B}_{\nu\lambda}=B_{\nu\lambda}+n\Bigl(\frac{\partial\psi_\nu}{\partial X^\lambda}-\psi_\nu\psi_\lambda\Bigr)-\Bigl(\frac{\partial\psi_\lambda}{\partial X^\nu}-\psi_\nu\psi_\lambda\Bigr)\nonumber\\
&&{\bar F}_{\lambda\nu}=F_{\lambda\nu}+\frac{n+1}{2}\Bigl(\frac{\partial\psi_\lambda}{\partial X^\nu}-\frac{\partial\psi_\nu}{\partial X^\lambda}\Bigr)
\end{eqnarray}
\noindent Eq.~(\ref{ts18}) shows that $F_{\lambda\nu}$ can be written as the curl of the vector $a_\nu/2$ defined as \cite{eisenhart} 
\begin{equation}\label{ts21}
a_\nu=\Gamma_{\kappa\nu}^\kappa-
\begin{Bmatrix} 
\kappa \\ \kappa\nu
\end{Bmatrix}
\end{equation}
\noindent Consequently, by choosing
\begin{equation}\label{ts22}
\psi_{\nu}=-\frac{1}{n+1}\Bigl(\Gamma_{\kappa\nu}^\kappa-
\begin{Bmatrix} 
\kappa \\ \kappa\nu
\end{Bmatrix}\Bigr)
\end{equation}
\noindent we have ${\bar F}_{\lambda\nu}=0$ and ${\bar R}_{\nu\lambda}={\bar B}_{\nu\lambda}$. From Eqs~(\ref{ts20}), we also have that the thermodynamic tensor ${\bar{\bar R}}_{\nu\lambda}$ remains symmetric for projective transformations of connection if, and only if, the projective covariant vector is the gradient of an arbitrary function of the $X$'s \cite{eisenhart}. In this case, the thermodynamic tensor ${\bar{\bar F}}_{\lambda\nu}$ remains unaltered i.e., ${\bar{\bar F}}_{\lambda\nu}=0$. Hence, at this stage, the expression of the affine connection is determined up to the gradient of a function, say $\phi$, of the thermodynamic forces, which is also {\it scalar under TCT}. Let us impose now the projective-invariance. Eqs~(\ref{ts20}) indicate that a necessary and sufficient condition that ${\bar R}_{\nu\lambda}$ be projective-invariant is that 
\begin{eqnarray}\label{ts23}
&&\!\!\!\!\!\!\!\!\!\!\!\!\!\!\!\!\!\!\!\!\!\!\!\!\!\!\!\!\!\!\!\!\!\!\!\!\!\!\!\!\!\!\!\!\!\!\!\!
\frac{\partial^2\phi}{\partial X^\lambda\partial X^\nu}-\frac{\partial\phi}{\partial X^\lambda}\frac{\partial\phi}{\partial X^\nu}=0\qquad{\rm with}\nonumber\\
&&\!\!\!\!\!\!\!\!\!\!\!\!\!\!\!\!\!\!\!\!\!\!\!\!\!\!\!\!\!\!\!\!\!\!\!\!\!\!\!\!\!\!\!\!\!\!\!\!
\left\{ \begin{array}{ll}
\phi=0 & \\
\frac{\partial\phi}{\partial X^\mu}=0 \\
\frac{\partial^2\phi}{\partial X^\mu\partial X^\nu}=0 \\
\end{array}
\right.\ \ ({\rm in\ the\ Onsager\ region})
\end{eqnarray}
\noindent where $\phi$ is a function, invariant under TCT. The solution of Eq.~(\ref{ts23}) is $\phi\equiv 0$ everywhere. The final expression of the affine connection for symmetric processes reads then
\begin{equation}\label{ts24}
\Gamma^\mu_{\alpha\beta}\!=\!
\begin{Bmatrix} 
\mu \\ \alpha\beta
\end{Bmatrix}
\!+\!\frac{1}{2\sigma}X^\mu\mathcal{O}(g_{\alpha\beta})\!-\!
\frac{1}{2(n+1)\sigma}\Bigl[
\delta^\mu_\alpha X^\nu\mathcal{O}(g_{\beta\nu})\!+\!\delta^\mu_\beta X^\nu\mathcal{O}(g_{\alpha\nu})\Bigr]
\end{equation}
\noindent The general case is given by
\begin{eqnarray}\label{ts25}
\ \!\Gamma^\mu_{\alpha\beta}=\!\!\!\!&&{\check N}^{\mu\kappa}g_{\kappa\lambda}\begin{Bmatrix} 
\lambda \\ \alpha\beta
\end{Bmatrix}+\frac{{\check N}^{\mu\kappa}}{2\sigma}X_\kappa\mathcal{O}(g_{\alpha\beta})
+\frac{{\check N}^{\mu\kappa} }{2\sigma}X_\kappa X^\lambda\Bigl(\frac{\partial f_{\alpha\lambda}}{\partial X^\beta}+\frac{\partial f_{\beta\lambda}}{\partial X^\alpha}\Bigr)\nonumber\\
&&+\frac{{\check N}^{\mu\kappa}}{2\sigma}f_{\kappa\varsigma}X^\varsigma X^\lambda\Bigl(\frac{\partial g_{\alpha\lambda}}{\partial X^\beta}+\frac{\partial g_{\beta\lambda}}{\partial X^\alpha}\Bigr)+\psi_\alpha\delta^\mu_\beta+\psi_\beta\delta^\mu_\alpha
\end{eqnarray}
\noindent 
where
\begin{eqnarray}\label{ts25a}
\psi_\nu=&&\!\!\!\!\!\!\!\!\!\!\!
-\frac{{\check N}^{\eta\kappa}g_{\kappa\lambda}}{n+1}\begin{Bmatrix} 
\lambda \\ \eta\nu
\end{Bmatrix}
\!-\frac{{\check N}^{\eta\kappa}X_\kappa}{2(n+1)\sigma}\mathcal{O}(g_{\nu\eta})\!-\!
\frac{{\check N}^{\eta\kappa} }{2(n+1)\sigma}X_\kappa X^\lambda\Bigl(\frac{\partial f_{\eta\lambda}}{\partial X^\nu}+\frac{\partial f_{\nu\lambda}}{\partial X^\eta}\Bigr)
\nonumber\\
&&\!\!\!\!\!\!\!\!\!\!\!
-\frac{{\check N}^{\eta\kappa}}{2(n+1)\sigma}f_{\kappa\varsigma}X^\varsigma X^\lambda\Bigl(\frac{\partial g_{\eta\lambda}}{\partial X^\nu}+\frac{\partial g_{\nu\lambda}}{\partial X^\eta}\Bigr)+\frac{1}{n+1}\frac{\partial\log\sqrt g}{\partial X^\nu}
\end{eqnarray}
Note that the thermodynamic space tends to reduce to a (thermodynamic) Riemannian space when ${\sigma}^{-1}\ll 1$. The following definitions are adopted:
\begin{itemize}
\item{ The space, covered by $n$ independent thermodynamic forces $X^\mu$, with metric tensor $g_{\mu\nu}$ and a linear single connection given by Eq.~(\ref{ts25}), may be referred to as {\it thermodynamic space} $T\!s$ (or, thermodynamical forces space).} 
\end{itemize}
In $T\!s$, the length of an arc is defined by the formula 
\begin{equation}\label{ts26}
L=\int_{\varsigma_1}^{\varsigma_2}\Bigl(g_{\mu\nu}\frac{ dX^\mu}{d\varsigma} \frac{ dX^\nu}{d\varsigma}\Bigr)^{1/2}d\varsigma
\end{equation}
\noindent The positive definiteness of matrix $g_{\mu\nu}$ ensures that $L\ge0$. Consider a coordinate system $X^\mu$, defining the thermodynamic space $T\!s$. 
\begin{itemize}
\item{All thermodynamic spaces obtained from $T\!s$ by a TCT transformation, may be called {\it entropy-covariant spaces}.} 
\end{itemize}
In the TFT description, a thermodynamic configuration corresponds to a point in the thermodynamic space $T\!s$. The equilibrium state is the origin of the axes. Consider a thermodynamic system out of equilibrium, represented by a certain point, say $a$, in the thermodynamic space
\begin{itemize}
\item{A thermodynamic system is said {\it to relax} (from the geometrical point of view) towards another point of the thermodynamic space, say $b$, if it moves from point $a$ to point $b$ following the shortest path~(\ref{ts3}), with the affine connection given in Eq.~(\ref{ts25}). Note that in this context the term relaxation refers to a relaxation in a geometrical sense.}
\end{itemize}
\begin{itemize}
\item{ With Eq.~(\ref{ts25}), Eqs~(\ref{ts6}) may be called the {\it thermodynamic covariant differentiation} of a thermodynamic vector while Eqs.~(\ref{ts2}) and (\ref{ts5}) the {\it thermodynamic covariant differentiation along a curve} of a thermodynamic vector.}
\end{itemize}
\begin{itemize}
\item{With affine connection Eq.~(\ref{ts25}), $R^{\mu}_{\nu\lambda\kappa}$ may be called the {\it thermodynamic curvature tensor}.} 
\end{itemize}
\begin{itemize}
\item{The scalar $R$ obtained by contracting the thermodynamic tensor $R_{\nu\lambda}$ with the symmetric piece of the transport coefficients (i.e. $R=R_{\nu\lambda}g^{\nu\lambda}$) may be called the {\it thermodynamic curvature scalar}.} 
\end{itemize}
\vskip 0.2truecm
\noindent{\bf The Principle of Least Action}
\vskip 0.2truecm
From expression (\ref{ts25}), the following mixed thermodynamic tensor of third order can be constructed
\begin{eqnarray}\label{pa1}
\Psi^\mu_{\alpha\beta}\equiv\!\!\!\!&&{\check N}^{\mu\kappa}g_{\kappa\lambda}\begin{Bmatrix} 
\lambda \\ \alpha\beta
\end{Bmatrix}+\frac{{\check N}^{\mu\kappa}}{2\sigma}X_\kappa\mathcal{O}(g_{\alpha\beta})
+\frac{{\check N}^{\mu\kappa} }{2\sigma}X_\kappa X^\lambda\Bigl(\frac{\partial f_{\alpha\lambda}}{\partial X^\beta}+\frac{\partial f_{\beta\lambda}}{\partial X^\alpha}\Bigr)\nonumber\\
&&+\frac{{\check N}^{\mu\kappa}}{2\sigma}f_{\kappa\varsigma}X^\varsigma X^\lambda\Bigl(\frac{\partial g_{\alpha\lambda}}{\partial X^\beta}+\frac{\partial g_{\beta\lambda}}{\partial X^\alpha}\Bigr)+\psi_\alpha\delta^\mu_\beta+\psi_\beta\delta^\mu_\alpha-\begin{Bmatrix} 
\mu \\ \alpha\beta
\end{Bmatrix}
\end{eqnarray}
\noindent This thermodynamic tensor satisfies the important identities
\begin{equation}\label{pa2}
\Psi^\alpha_{\alpha\beta}=\Psi^\beta_{\alpha\beta}=0
\end{equation}
\noindent Again, from $\Psi^\mu_{\alpha\beta}$ the mixed thermodynamic tensor of fifth order can be constructed
\begin{equation}\label{pa3}
S^{\mu\nu}_{\lambda\alpha\beta}\equiv 
\frac{1}{2}\Bigl(\Psi^\mu_{\beta\lambda}\delta^\nu_{\alpha}+
\Psi^\mu_{\alpha\lambda}\delta^\nu_{\beta}+
\Psi^\nu_{\beta\lambda}\delta^\mu_{\alpha}+
\Psi^\nu_{\alpha\lambda}\delta^\mu_{\beta}
-\Psi^\mu_{\alpha\beta}\delta^\nu_\lambda
-\Psi^\nu_{\alpha\beta}\delta^\mu_\lambda\Bigr)
\end{equation}
\noindent By contraction, a thermodynamic tensor of third order, a thermodynamic vector and a thermodynamic scalar can be formed as follows
\begin{eqnarray}\label{pa4}
&&\!\!\!\!\!\!\!\!\!\!\!\!\!\!\!\!\!\!\!\!
S^{\mu\nu}_{\lambda}\equiv S^{\mu\nu}_{\lambda\alpha\beta}g^{\alpha\beta}=\Psi^\mu_{\lambda\alpha}g^{\nu\alpha}+
\Psi^\nu_{\lambda\alpha}g^{\mu\alpha}-\frac{1}{2}\Psi^\mu_{\alpha\beta}g^{\alpha\beta}\delta^\nu_\lambda-\frac{1}{2}\Psi^\nu_{\alpha\beta}g^{\alpha\beta}\delta^\mu_\lambda\nonumber\\
&&\!\!\!\!\!\!\!\!\!\!\!\!\!\!\!\!\!\!\!\!
S^{\mu}\equiv S^{\mu\lambda}_{\lambda}=\frac{1-n}{2}\Psi^\mu_{\alpha\beta}g^{\alpha\beta}\\
&&\!\!\!\!\!\!\!\!\!\!\!\!\!\!\!\!\!\!\!\!
S\equiv S^{\mu\nu}_{\lambda}\Psi^{\lambda}_{\mu\nu}=2\Psi^\kappa_{\lambda\mu}\Psi^\lambda_{\kappa\nu}g^{\mu\nu}\nonumber
\end{eqnarray}
\noindent The following postulate is now introduced:

\noindent {\it There exists a thermodynamic action $I$, scalar under $TCT$, which is stationary with respect to arbitrary variations in the transport coefficients and the affine connection}. 

\noindent This action, scalar under $TCT$, is constructed from the transport coefficients, the affine connection and their first derivatives. In addition, it should have linear second derivatives of the transport coefficients and it should not contain second (or higher) derivatives of the affine connection. We also require that the action is stationary when the affine connection takes the expression given in Eq.~(\ref{ts25}). The only action satisfying these requirements is 
\begin{equation}\label{pa5}
I=\int\Bigl[ R_{\mu\nu}-(\Gamma^\lambda_{\alpha\beta}-
{\tilde\Gamma}^\lambda_{\alpha\beta})S^{\alpha\beta}_{\lambda\mu\nu}
\Bigr]g^{\mu\nu}\sqrt{g}\ \! {d^{}}^n\!X
\end{equation}
\noindent where ${d^{}}^n\!X$ denotes an infinitesimal volume element in $T\!s$ and ${\tilde\Gamma}^\kappa_{\mu\nu}$ is the expression given in Eq.~(\ref{ts25}) i.e., ${\tilde\Gamma}^\kappa_{\mu\nu}=\Psi^\kappa_{\mu\nu}+\Bigl\{{}^{\ \!\kappa}_{\mu\nu}\Bigr\}$. To avoid misunderstanding, while it is correct to mention that this postulate affirms the possibility of deriving the nonlinear closure equations by a variational principle it does not state that the expressions and theorems obtained from the solutions of these equations can also be derived by a variational principle. In particular the well-known {\it Universal Criterion of Evolution} established by Glansdorff-Prigogine {\it can not} be derived by a variational principle (see also section \ref{ttse}).
\vskip 0.2truecm
\noindent{\bf The Nonlinear Closure Equations}
\vskip 0.2truecm
The transport coefficients and the affine connection should be considered as independent dynamical variables (as opposed to $X^\mu$, which is a mere variable of integration) \cite{palatini}. Therefore, the action (\ref{pa5}) is stationary with respect to arbitrary variations in $g_{\mu\nu}$, $f_{\mu\nu}$ and $\Gamma^\lambda_{\mu\nu}$. As a first step, we suppose that the transport coefficients and the affine connection be subject to infinitesimal variations i.e., $g_{\mu\nu}\rightarrow g_{\mu\nu}+\delta g_{\mu\nu}$, $f_{\mu\nu}\rightarrow f_{\mu\nu}+\delta f_{\mu\nu}$ and $\Gamma^\kappa_{\mu\nu}\rightarrow \Gamma^\kappa_{\mu\nu}+\delta \Gamma^\kappa_{\mu\nu}$, where $\delta g_{\mu\nu}$, $\delta f_{\mu\nu}$ and $\delta \Gamma^\kappa_{\mu\nu}$ are arbitrary, except that they are required to vanish as $\mid X^\mu\mid\rightarrow\infty$. Upon application of the principle of stationary action, the following {\it nonlinear closure equations} (i.e., the equations for the transport coefficients and the affine connection) are derived (see appendix \ref{appeq}):
\begin{eqnarray}\label{tfe1}
&& R_{\mu\nu}-\frac{1}{2}g_{\mu\nu}R=-S^{\alpha\beta}_\lambda\frac{\delta{\tilde\Gamma}^\lambda_{\alpha\beta}}{\delta g^{\mu\nu}}\nonumber\\
&&S^{\alpha\beta}_\lambda\frac{\delta{\tilde\Gamma}^\lambda_{\alpha\beta}}{\delta f^{\mu\nu}}=0\\
&&g_{\mu\nu\mid\lambda}=-\Psi^\alpha_{\mu\lambda}g_{\alpha\nu}-\Psi^\alpha_{\nu\lambda}g_{\alpha\mu}\nonumber
\end{eqnarray}
\noindent where the variations of the affine connection (\ref{ts25}) with respect to the transport coefficients appear in the first two equations. Notice that $R_{\mu\nu}-\frac{1}{2}g_{\mu\nu}R$ {\it does not} coincide with Einstein's tensor (see also Appendix \ref{comparison}). From the first equation of Eqs~(\ref{tfe1}), and for $n\neq 2$, the expression for the thermodynamic curvature scalar is obtained \footnote{Eq.~(\ref{tfe2a}) does not apply in two dimensions. Two-dimensional problem may be met in the limit case of a system driven out of equilibrium by two (independent) {\it scalar} thermodynamic forces such as, for example, two chemical affinities (and {\it not} when we analyse, for example, a system submitted to two vectorial thermodynamic forces in one-dimension), where diffusion of the chemical species is neglected. This ideal example is, however, analyzed in ref. \cite{sonnino4}. Eq.~(\ref{tfe2a}) should be replaced by $R=2R_{1212}/g$ where $R_{1212}$ and $g$ indicate the $1212$ component of the thermodynamic curvature tensor and the determinant of the matrix $g_{\mu\nu}$, respectively (see, for example, ref. \cite{weinberg}).\label{space}}
\begin{equation}\label{tfe2a}
R=\frac{2}{n-2}g^{\mu\nu}S^{\alpha\beta}_\lambda \frac{\delta{\tilde\Gamma}^\lambda_{\alpha\beta}}{\delta g^{\mu\nu}}\qquad\quad (n\neq 2)
\end{equation}
\noindent The third equation of Eqs~(\ref{tfe1}) can be re-written as
\begin{equation}\label{tfe2b}
g_{\mu\nu,\lambda}
-\Gamma^\alpha_{\mu\lambda}g_{\alpha\nu}-\Gamma^\alpha_{\nu\lambda}g_{\alpha\mu}
=-\Psi^\alpha_{\mu\lambda}g_{\alpha\nu}-\Psi^\alpha_{\nu\lambda}g_{\alpha\mu}
\end{equation}
\noindent where the comma $(,)$ denotes partial differentiation. Adding to this equation the same equation with $\mu$ and $\lambda$ interchanged and subtracting the same equation with $\nu$ and $\lambda$ interchanged gives
\begin{equation}\label{tfe2c}
g_{\mu\nu,\lambda}+g_{\lambda\nu,\mu}-g_{\mu\lambda,\nu}=2g_{\alpha\nu}\Gamma^\alpha_{\lambda\mu}-2g_{\alpha\nu}\Psi^\alpha_{\lambda\mu}
\end{equation}
\noindent or 
\begin{equation}\label{tfe3}
\Gamma^\kappa_{\lambda\mu}=\Bigl\{{}^{\ \!\kappa}_{\lambda\mu}\Bigr\}+\Psi^\kappa_{\lambda\mu}={\tilde\Gamma}^\kappa_{\lambda\mu}
\end{equation}
\noindent Hence, action Eq.~(\ref{pa5}) is stationary when the affine connection takes the expression given in Eq.~(\ref{ts25}). For $a-a$ or $b-b$ processes, close to the Onsager region, it holds that
\begin{eqnarray}\label{tfe5}
\!\!\!\!\!\!\!\!\!\!\!\!\!\!\!\!&&g_{\mu\nu}=L_{\mu\nu}+h_{\mu\nu}+O(\epsilon^2)\nonumber\\
\!\!\!\!\!\!\!\!\!\!\!\!\!\!\!\!&&\lambda_\sigma=O(\epsilon)\qquad {\rm with}\quad \epsilon=Max\Big\{\frac{\mid {\rm Eigenvalues} [g_{\mu\nu}-L_{\mu\nu}]\mid}{{\rm Eigenvalues} [L_{\mu\nu}]}\Big\}\ll 1
\end{eqnarray}
\noindent where $\lambda_\sigma\equiv 1/\sigma$ and $h_{\mu\nu}$ are small variations with respect to Onsager's coefficients. In this region, Eq.~(\ref{pa5}) is stationary for arbitrary variations of $h_{\mu\nu}$ and $\Gamma^\kappa_{\mu\nu}$. It can be shown that \cite{sonnino}
\begin{eqnarray}\label{tfe6}
\!\!\!\!\!\!\!\!\!\!\!\!&&L^{\lambda\kappa}\frac{\partial^2 h_{\mu\nu}}{\partial X^{\lambda}\partial X^{\kappa}}+
L^{\lambda\kappa}\frac{\partial^2 h_{\lambda\kappa}}{\partial X^{\mu}\partial X^{\nu}}-
L^{\lambda\kappa}\frac{\partial^2 h_{\lambda\nu}}{\partial X^{\kappa}\partial X^{\mu}}-
L^{\lambda\kappa}\frac{\partial^2 h_{\lambda\mu}}{\partial X^{\kappa}\partial X^{\nu}}=0+O(\epsilon^2)\nonumber\\
\!\!\!\!\!\!\!\!\!\!\!\!&&\Gamma_{\mu\nu}^\kappa=\frac{1}{2}L^{\kappa\eta}(h_{\mu\eta,\nu}+h_{\nu\eta,\mu}-h_{\mu\nu,\eta})+O(\epsilon^2)
\end{eqnarray}
\noindent Eqs~(\ref{tfe6}) should be solved with the appropriate gauge-choice and boundary conditions.

\noindent The validity of Eqs~(\ref{tfe6}) has been largely tested by analyzing several symmetric processes, such as the thermoelectric effect and the unimolecular triangular chemical reactions \cite{sonnino}. More recently, these equations have been also used to study transport processes in magnetically confined plasmas. In all examined examples, the theoretical results of the TFT are in line with experiments. It is worthwhile mentioning that, for transport processes in tokamak plasmas, the predictions of the TFT for radial energy and matter fluxes are much closer to the experimental data than the neoclassical theory, which fails with a factor $10^3\div10^4$ \cite{sonnino3}, \cite{balescu2}. The physical origin of this failure can be easily understood. As mentioned in the introduction, even in absence of turbulence, the state of the plasma is close to, but not in, a state of local equilibrium. Indeed, starting from an arbitrary initial state, the collisions would tend, if they were alone, to bring the system very quickly to a local equilibrium state. But slow processes, i.e. free-flow and electromagnetic processes, prevent the plasma from reaching this state. The distribution function for the fluctuations of the thermodynamic quantities also deviates from a Maxwellian preventing the thermodynamic fluxes from being linearly connected with the conjugate forces (ref. to the Onsager theory \cite{onsager} and, for example, \cite{degroot}). In tokamak plasmas, the thermodynamic forces and the conjugate flows are the generalized frictions and the Hermitian moments, respectively \cite{balescu2}. In the neoclassical theory, the flux-force relations have been truncated at the linear order (ref., for example, to \cite{balescu1}), in contrast with the fact that the distribution function of the thermodynamic fluctuations is not a Maxwellian. This may be one of the main causes of the strong disagreement between the neoclassical previsions and the experimental profiles \footnote{For more details, ref. to G. Sonnino {\it Some Considerations on the Basic Assumptions of the Neoclassical Theory}, to be sent to {\it Physical Review Letters} (2009). \label{neocl}} \cite{sonnino3}. It is, however, important to mention that it is well accepted that another main reason of this discrepancy is attributed to turbulent phenomena existing in tokamak plasmas. Fluctuations in plasmas can become unstable and therefore amplified, with their nonlinear interaction successively leading the plasma to a state, which is far away from equilibrium. In this condition, the transport properties are supposed to change significantly and to exhibit qualitative features and properties that could not be explained by collisional transport processes, e.g. size-scaling with machine dimensions and non-local behaviors that clearly point at turbulence spreading etc. The scope of the work cited in ref.\cite{sonnino3} is mainly to demonstrate that collisional transport processes in fusion plasmas can be computed via a nonlinear theory on a more rigorous and sound basis than that provided by the well known classical and/or neoclassical theory. The proposed approach includes prior known results as a limiting case where nonlinear and non-local effects in collisional transport processes can be ignored. More generally, the TFT estimates of collisional transport fluxes can be amplified by up to two or three orders of magnitude with respect to the classical/neoclassical levels in the electron transport channel, while ions corrections are much smaller. However, TFT collisional transport levels remain a fraction of the values observed experimentally, confirming that turbulent transport is the generally dominant process determining particle and heat fluxes in magnetically confined plasmas. In this specific example, the nonlinear corrections provide with an evaluation of the (parallel) Hermitian moments of the electron and ion distribution functions \cite{sonnino3}. 

\vskip 0.2truecm
\noindent{\bf Some Remarks on Spatially-Extended Thermodynamic Systems}
\vskip 0.2truecm
The macroscopic description of thermodynamic systems gives rise to state variables that depend continuously on space coordinates. In this case, the thermodynamic forces possess an infinity associated to each point of the space coordinates. The system may be subdivided into $N$ cells ($N$x$N$x$N$ in three dimensions), each of which labeled by a wave-number ${\bf k}$, and we follow their relaxation. Without loss of generality, we consider a thermodynamic system confined in a rectangular box with sizes $l_x$, $l_y$ and $l_z$. We write the wave-number as 
\begin{equation}\label{sets1}
{\bf k}=2\pi\bigl(\frac{n_x}{l_x}, \frac{n_y}{l_y}, \frac{n_z}{l_z}\bigr)\quad{\rm with}\quad
\left\{ \begin{array}{ll}
n_x= 0,\pm1,\cdots\pm N_x\\
n_y= 0,\pm1,\cdots\pm N_y\\
n_z= 0,\pm1,\cdots\pm N_z
\end{array}
\right.
\end{equation}
\noindent The fluxes and forces, developed in (spatial) Fourier's series, read
\begin{eqnarray}\label{sets2}
&&\!\!\!\!\!\!\!\!\!\!\!\!
{\mathcal J}_\mu({\bf r},t)=\!\!\!\!\sum_{{\bf n}=-{\bf N}}^{\bf N}{\hat J}_{\mu ({\bf k})}(t)\exp(i{\bf k}\cdot{\bf r})\nonumber\\
&&\!\!\!\!\!\!\!\!\!\!\!\!
{\mathcal X}^\mu({\bf r},t)=\!\!\!\!\sum_{{\bf n}'=-{\bf N}}^{\bf N}{\hat X}^\mu_{({\bf k}')}(t)\exp(i{\bf k}'\cdot{\bf r})
\end{eqnarray}
\noindent where, for brevity, ${\bf n}$ and ${\bf N}$ stand for ${\bf n}=(n_x,n_y,n_z)$ and ${\bf N}=(N_x,N_y,N_z)$, respectively. The Fourier coefficients are given by 
\begin{eqnarray}\label{sets3}
&&\!\!\!\!\!\!\!\!\!\!\!\!
{\hat J}_{\mu ({\bf k})}(t)=\frac{1}{\Omega}\int_\Omega {\mathcal J}_\mu({\bf r},t)
\exp(-i{\bf k}\cdot{\bf r})dv\nonumber\\
&&\!\!\!\!\!\!\!\!\!\!\!\!
{\hat X}^\mu_{({\bf k}')}(t)=\frac{1}{\Omega}\int_\Omega{\mathcal X}^\mu({\bf r},t)
\exp(-i{\bf k}'\cdot{\bf r})dv
\end{eqnarray}
\noindent In particular, the contributions at the thermodynamic limit (i.e., for ${\bf k}\rightarrow 0$) are expressed as 
\begin{eqnarray}\label{sets4}
&&\!\!\!\!\!\!\!\!\!\!\!\!
{\hat J}_{\mu (0)}(t)=\frac{1}{\Omega}\int_\Omega {\mathcal J}_\mu({\bf r},t)
dv= J_\mu(t)\nonumber\\
&&\!\!\!\!\!\!\!\!\!\!\!\!
{\hat X}^\mu_{(0)}(t)=\frac{1}{\Omega}\int_\Omega{\mathcal X}^\mu({\bf r},t)
dv= X^\mu(t)
\end{eqnarray}
\noindent The entropy production and the fluxes-forces relations take, respectively, the form
\begin{eqnarray}\label{sets5}
&&\!\!\!\!\!\!\!\!\!\!\!\!
 {\mathcal\sigma}({\bf r},t)={\mathcal J}_\mu({\bf r},t){\mathcal X}^\mu({\bf r},t)\geq0
\nonumber\\
&&\!\!\!\!\!\!\!\!\!\!\!\!
{\mathcal J}_\mu({\bf r},t)={\mathcal\tau}_{\mu\nu}({\bf r},t){\mathcal X}^\nu({\bf r},t)
\end{eqnarray}
\noindent Considering that 
\begin{eqnarray}\label{sets6}
&&\!\!\!\!\!\!\!\!\!\!\!\!\!\!\!\!\!\!\!\!\!\!\!\!\!\!\!\!\!\!\!\!\!\!\!\!
\int_0^{l_x}\!\!\int_0^{l_y}\!\!\int_0^{l_z}\!\!\exp[i({\bf k}+{\bf k}')\cdot{\bf r}]dv =\Omega\ \delta_{{\bf k}+{\bf k}',0}\qquad {\rm with}\\
&&\!\!\!\!\!\!\!\!\!\!\!\!\!\!\!\!\!\!\!\!\!\!\!\!\!\!\!\!\!\!\!\!\!\!\!\!
\delta_{{\bf k}+{\bf k}',0}=
\left\{ \begin{array}{ll}
0& \mbox{if $ {\bf k}+{\bf k}'\neq 0$}\\
1& \mbox{if $ {\bf k}+{\bf k}'= 0$}
\end{array}\qquad\qquad {\rm and}\qquad\!\! \Omega=l_xl_yl_z
\right.\nonumber
\end{eqnarray}
\noindent from the first equation of Eq.~(\ref{sets5}) we also find
\begin{equation}\label{sets7}
\int_\Omega
{\mathcal J}_\mu({\bf r},t){\mathcal X}^\mu({\bf r},t)\ dv=\Omega\Bigl({\hat J}_{\mu (0)}(t){\hat X}^\mu_{(0)}(t)+
\sum_{{\bf k}\neq0}
{\hat J}_{\mu ({\bf k})}(t){\hat  X}^\mu_{({\bf -k})}(t)\Bigr)\geq0
\end{equation}
\noindent On the other hand, we have
\begin{equation}\label{sets8}
{\hat J}_{\mu (0)}(t)={\hat \tau}_{\mu\nu(0)}(t){\hat X}^\nu_{(0)}(t)+\sum_{{\bf k}\neq0}{\hat \tau}_{\mu\nu({\bf k})}(t){\hat X}^\nu_{({\bf -k})}(t)
\end{equation}
\noindent where
\begin{equation}\label{sets7a}
{\hat \tau}_{\mu\nu({\bf k})}(t)=\frac{1}{\Omega}\int_\Omega \tau_{\mu\nu}({\bf r},t)
\exp(-i{\bf k}\cdot{\bf r})dv
\end{equation}
\noindent Eq.~(\ref{sets7}) can then be brought into the form
\begin{eqnarray}\label{sets9}
&&\!\!\!\!\!\!\!\!\!\!\!\!\!\!\!\!\!\!\!
\int_\Omega
\sigma\ dv=\Omega{\hat g}_{\mu\nu(0)}(t){\hat X}^\mu_{(0)}(t){\hat X}^\nu_{(0)}(t)\nonumber\\
&&\quad\ 
+\Omega
\sum_{{\bf k}\neq0}\Bigl({\hat \tau}_{\mu\nu({\bf k})}(t){\hat X}^\nu_{({\bf -k})}(t){\hat X}^\nu_{(0)}(t)
+{\hat J}_{\mu ({\bf k})}(t){\hat  X}^\mu_{({\bf -k})}(t)\Bigr)\geq0
\end{eqnarray}
\noindent where
\begin{eqnarray}\label{sets9a}
&&{\hat g}_{\mu\nu({\bf k})}(t)=\frac{1}{\Omega}\int_\Omega {\mathcal G}_{\mu\nu}({\bf r},t)
\exp(-i{\bf k}\cdot{\bf r})dv\qquad{\rm with}\nonumber\\
&&{\mathcal G}_{\mu\nu}({\bf r},t)\equiv\frac{1}{2}[\tau_{\mu\nu}({\bf r},t)+\tau_{\nu\mu}({\bf r},t)]
\end{eqnarray}
\noindent In Eq.~(\ref{sets9}), the first term is the contribution at the thermodynamic limit whereas the second expression reflects the interactions between the ${\bf k}$-cell and the other cells. In a relaxation process, contributions from different wave-numbers are negligible with respect to those with same wave-numbers ({\it the slaving principle} \cite{haken}) and, hence, we finally obtain 
\begin{equation}\label{sets10}
\int_\Omega
\sigma\ dv\simeq\Omega
{\hat g}_{\mu\nu(0)}(t){\hat X}^\mu_{(0)}(t){\hat X}^\nu_{(0)}(t)>0\qquad\forall\ {\hat X}^\mu_{(0)}(t)\ ({\rm and}\ \sigma\neq0)
\end{equation}
\noindent Last inequality is satisfied {\it for any} ${\hat X}^\mu_{(0)}(t)$ if, and only if
\begin{equation}\label{sets11}
{\hat g}_{\mu\nu(0)}(t)=\frac{1}{\Omega}\int_\Omega{\mathcal G}_{\mu\nu}({\bf r},t)\ dv=g_{\mu\nu}(t)
\end{equation}
\noindent is a {\it positive definite matrix}. This non-trivial result will be extensively used in Section \ref{ttse}. For spatially-extended thermodynamic systems, we have then to replace $X^{\mu}(t)\rightarrow X^\mu_{({\bf k})}(t)$ and $\tau_{\mu\nu} (t)\rightarrow \tau_{\mu\nu({\bf k})} (t)$. Under these conditions, Eqs~(\ref{tfe1}) determine the nonlinear corrections to the Onsager coefficients while Eqs~(\ref{ts2}) and Eqs~(\ref{ts6}), with affine connection Eq.~(\ref{ts25}), are the thermodynamic covariant differentiation along a curve and the thermodynamic covariant differentiation of a thermodynamic vector, respectively. 

\vskip 0.2truecm
\noindent{\bf The Privileged Thermodynamic Coordinate System}
\vskip 0.2truecm
By definition, a thermodynamic coordinate system is a set of coordinates defined so that the expression of the entropy production takes the form of Eq.~(\ref{tr1}). Once a particular set of thermodynamic coordinates is determined, the other sets of coordinates are linked to the first one through a TCT [see Eqs~(\ref{tr2})]. The simplest way to determine a particular set of coordinates is to quote the entropy balance equation
\begin{equation}\label{ptcs1}
\frac{\partial \rho s}{\partial t}+{\bf\nabla}\cdot{\bf J}_s=\sigma
\end{equation}
 \noindent where $\rho s$ is the local total entropy per unit volume, and ${\bf J}_s$ is the entropy flux. Let us consider, as an example, a thermodynamic system confined in a rectangular box where chemical reactions, diffusion of matter, macroscopic motion of the volume element (convection) and heat current take place simultaneously. The entropy flux and the entropy production read \cite{prigogine3}, \cite{fitts}
 \begin{eqnarray}\label{pct2}
&&\!\!\!\!\!\!\!\!\!\!\!\!\!\!\!\!\!\!\!\!
 {\bf J}_s=\frac{1}{T}({\bf J}_q-\sum_i {\bf J}_i\mu_i)+\sum_i\rho_i v_is_i\nonumber\\
&&\!\!\!\!\!\!\!\!\!\!\!\!\!\!\!\!\!\!\!\!
\sigma={\bf J}_q\!\cdot\!{\bf\nabla}\frac{1}{T}\!-\!\frac{1}{T}\!\sum_i{\bf J}_i\!\cdot\!\Bigl[T{\bf\nabla}\Bigl(\frac{\mu_i}{T}\Bigl)\!-\!{\bf F}_i\Bigr]\!+\!\sum_i\frac{w_iA_i}{T}\!-\!\frac{1}{T}\!\sum_{ij}\Pi_{ij}\partial_{{\bf r}_i}v_j\geq0\\
\end{eqnarray}
 \noindent where $\mu_i$, $\rho_i s_i$ and $A_i$ are the chemical potential, the local entropy and the affinity of species $"i"$, respectively. ${\bf J}_q$ is the heat flux; ${\bf J}_i$ and $w_i$ are the diffusion flux and the chemical reaction rate of species $"i"$, respectively. Moreover, $\Pi_{ij}$ indicate the components of the dissipative part of the pressure tensor ${\mathcal M}_{ij}$ (${\mathcal M}_{ij}=p\delta_{ij}+\Pi_{ij}$; $p$ is the hydrostatic pressure), ${\bf F}_i$ the external force per unit mass acting on $"i"$, and $v_j$ is the component of the hydrodynamic velocity (see, for example, ref. \cite{vidal}). The set of thermodynamic coordinates is given as
 \begin{equation}\label{ptcs3}
\Bigl\{{\bf\nabla}\frac{1}{T};\ -\frac{1}{T}\Bigl[T{\bf\nabla}\Bigl(\frac{\mu_i}{T}\Bigl)-{\bf F}_i\Bigr];\ \frac{A_i}{T};\ -\frac{1}{T}\partial_{{\bf r}_i}v_j\Bigr\}
\end{equation}
 \noindent For this particular example, this set may be referred to as the {\it privileged thermodynamic coordinate system}. Other examples of privileged thermodynamic coordinate system, concerning magnetically confined plasmas, can be found in refs \cite{sonnino3}, \cite{balescu2} and \cite{balescu1}.
 
 \vskip 0.2truecm
\noindent{\bf A Special Class of TCT: The Linear Transformations}
\vskip 0.2truecm

A case frequently encountered in literature occurs when, in all thermodynamic space, we perform the linear transformations 
\begin{eqnarray}\label{trl1}
&&X'^\mu=c^\mu_\nu X^\nu\nonumber\\
&& J'_\mu={\tilde c}^\nu_\mu J_\nu\qquad{\rm with}\qquad  {\tilde c}^\mu_\kappa c^\kappa_\nu=\delta^\mu_\nu
\end{eqnarray}
\noindent where $c^\mu_\nu$ is a constant matrix (i.e., independent of the thermodynamic forces). It can be checked that for this particular choice, we have the following [see also Appendix (\ref{appx})]
\vskip0.3cm
\begin{itemize}
\item{
The affine connection notably simplifies}
 \end{itemize}
\begin{equation}\label{trl2}
\Gamma^\mu_{\alpha\beta}=\!{\check N}^{\mu\kappa}g_{\kappa\lambda}\begin{Bmatrix} 
\lambda \\ \alpha\beta
\end{Bmatrix}\!+\!\frac{{\check N}^{\mu\kappa}}{2\sigma}X_\kappa\mathcal{O}(g_{\alpha\beta})
\!+\!\frac{{\check N}^{\mu\kappa} }{2\sigma}X_\kappa X^\lambda\Bigl(\frac{\partial f_{\alpha\lambda}}{\partial X^\beta}\!+\!\frac{\partial f_{\beta\lambda}}{\partial X^\alpha}\Bigr)
\!+\!\psi_\alpha\delta^\mu_\beta\!+\!\psi_\beta\delta^\mu_\alpha
\end{equation}
\noindent 
$\quad\quad\ \ $where
\begin{equation}\label{trl3}
\psi_\nu\!=\!\!
-\frac{{\check N}^{\eta\kappa}g_{\kappa\lambda}}{n+1}\begin{Bmatrix} 
\lambda \\ \eta\nu
\end{Bmatrix}
\!-\frac{{\check N}^{\eta\kappa}X_\kappa}{2(n+1)\sigma}\mathcal{O}(g_{\nu\eta})-
\frac{{\check N}^{\eta\kappa} }{2(n+1)\sigma}X_\kappa X^\lambda\Bigl(\frac{\partial f_{\eta\lambda}}{\partial X^\nu}+\frac{\partial f_{\nu\lambda}}{\partial X^\eta}\Bigr)
+\frac{1}{n+1}\frac{\partial\log\sqrt g}{\partial X^\nu}\nonumber
\end{equation}
\begin{itemize}
\item{
The balance equations for the thermodynamic forces (as well as the closure equations) are also covariant under TCT.}
\item{The nonlinear closure equations are given by Eqs~(\ref{tfe1}) with ${\tilde\Gamma}^\lambda_{\alpha\beta}$ provided by Eqs~(\ref{trl2}).}
\end{itemize}
\vskip0.3cm
\noindent Many examples of systems, analyzed by performing the linear transformations~(\ref{trl1}), can be found in ref.~\cite{degroot}.
\vskip 0.2truecm
\section{Thermodynamic Theorems for Systems out of Equilibrium}\label{ttse}
\vskip 0.2truecm
In 1947, Prigogine proved the minimum entropy production theorem \cite{prigogine}, which concerns the relaxation of thermodynamic systems near equilibrium. This theorem states that:
\vskip 0.2truecm
\noindent {\bf Minimum Entropy Production Theorem} (MEPT)

{\it For $a-a$ or $b-b$ processes, a thermodynamic system, near equilibrium, relaxes to a steady-state $X_s$ in such a way that the inequality}
\begin{equation}\label{tt1}
\frac{d\sigma}{dt}\le0
\end{equation}
\noindent {\it is satisfied throughout the evolution and is only saturated at $X_s$}.

\noindent The minimum entropy production theorem is generally not satisfied far from equilibrium. Indeed, under TCT, the rate of the entropy production transforms as
\begin{equation}\label{tt1a}
\frac{d\sigma'}{dt}=\frac{d\sigma}{dt}+\frac{\partial X'^\kappa}{\partial X^\eta}\frac{\partial^2X^\mu}{\partial  X^\nu\partial X'^\kappa}X^\eta J_\mu\frac{dX^\nu}{dt}
\end{equation}
\noindent In particular, we find
\begin{eqnarray}\label{tt1b}
&&\!\!\!\!\!\!\!\!\!\!\!\!\!\!\!\!J'_\mu\frac{dX'^\mu}{dt}=J_\mu\frac{dX^\mu}{dt}\nonumber\\
&&\!\!\!\!\!\!\!\!\!\!\!\!\!\!\!\!X'^\mu\frac{dJ'_\mu}{dt}=X^\mu\frac{dJ_\mu}{dt}+\frac{\partial X'^\kappa}{\partial X^\eta}\frac{\partial^2X^\mu}{\partial X^\nu\partial X'^\kappa}X^\eta J_\mu\frac{dX^\nu}{dt}
\end{eqnarray}
\noindent The second expression of Eqs~(\ref{tt1b}) tells us that nothing can be said about the sign of $X^\mu\frac{dJ_\mu}{dt}$. Concerning the quantity $J_\mu\frac{dX^\mu}{dt}$, Glansdorff and Prigogine \cite{prigogine1} demonstrated in 1954 a theorem, which reads
\vskip 0.2truecm
\noindent {\bf Universal Criterion of Evolution} (UCE)

{\it When the thermodynamic forces and conjugate flows are related by a generic asymmetric tensor, regardless of the type of processes, for time-independent boundary conditions a thermodynamic system, even in strong non-equilibrium conditions, relaxes towards a steady-state in such a way that the following universal criterion of evolution is satisfied}:
\begin{equation}\label{tt2}
{\mathcal P}\equiv J_\mu\frac{dX^\mu}{dt}\le0
\end{equation}
\noindent {\it This inequality is only saturated at $X_s$}. 

\noindent For $a-a$ or $b-b$ processes, the UCE reduces to the MEPT in the Onsager region. As mentioned in the introduction of this manuscript, Glansdorff and Prigogine demonstrated this theorem using a purely thermodynamical approach. In this section we shall see that if the system relaxes towards a steady-state along the shortest path then the Universal Criterion of Evolution is automatically satisfied. 

\noindent By definition, a necessary and sufficient condition for a curve to be the shortest path is that it satisfies the differential equation 
\begin{equation}\label{tt2a}
\frac{d^2X^\mu}{dt^2}+\Gamma^\mu_{\alpha\beta}\frac{dX^\alpha}{dt} \frac{dX^\beta}{dt}=\varphi(t)\frac{dX^\mu}{dt}
\end{equation}
\noindent where $\varphi(t)$ is a determined function of time. If we define a parameter $\varrho$ by
\begin{equation}\label{tt2b}
\frac{d\varrho}{dt}=c\exp\int\varphi^{*} dt\qquad{\rm with}\quad \varphi^{*}=\varphi-2\psi_\nu\frac{dX^\nu}{dt}
\end{equation}
\noindent where $c$ is an arbitrary constant and $\psi_\nu$ the projective covariant vector, Eq.~(\ref{tt2a}) reduces to Eq.~(\ref{ts3}) with $\Gamma^\mu_{\alpha\beta}$ given by Eq.~(\ref{ts9}). Parameter $\varrho$ is not the affine parameter $s$ of the shortest path. The relation between these two parameters is 
\begin{equation}\label{tt2c}
\varrho=b\int\exp\Bigl(-2\int\psi_\nu dX^\nu\Bigr)ds
\end{equation}
\noindent where $b$ is an arbitrary constant. Eq.~(\ref{tt2b}) allows us to choose the parameter $\varrho$ in such a way that it increases monotonically as the thermodynamic system evolves in time. In this case, $c$ is a positive constant and, without loss of generality, we can set $c=1$. Parameter $\varrho$ can also be chosen so that it vanishes when the thermodynamic system begins to evolve and it takes the (positive) value, say ${\bar l}$, when the system reaches the steady-state. Multiplying Eq.~(\ref{ts3}) with the flows $J_\mu$ and contracting, we obtain
\begin{equation}\label{tt3}
J_\mu\frac{d^2X^\mu}{d\varrho^2}+J_{\mu}\Gamma^\mu_{\alpha\beta}\frac{dX^\alpha}{d\varrho} \frac{dX^\beta}{d\varrho} =0
\end{equation}
\noindent However 
\begin{equation}\label{tt4}
 J_\mu\frac{d^2X^\mu}{d\varrho^2}=\frac{d{\tilde P}}{d\varrho}-\Bigl(\frac{d\varsigma}{d\varrho}\Bigr)^2-
 \frac{dX^\alpha}{d\varrho} \frac{dX^\beta}{d\varrho} 
  X^\lambda\frac{\partial g_{\alpha\lambda}}{\partial X^\beta}- \frac{dX^\alpha}{d\varrho} \frac{dX^\beta}{d\varrho}X^\lambda\frac{\partial f_{\alpha\lambda}}{\partial X^\beta}
\end{equation}
\noindent where ${\tilde P}=J_\mu \frac{dX^\mu}{d\varrho} $. In Eq.~(\ref{tt4}) the identities $f_{\mu\nu}\frac{dX^\mu}{d\varrho} \frac{dX^\nu}{d\varrho} =0$ and $g_{\mu\nu}\frac{dX^\mu}{d\varsigma} \frac{dX^\nu}{d\varsigma} =1$ have been taken into account. In addition, recalling Eq.~(\ref{ts13}) and the relations $X_\mu X^\mu=\sigma$ and $f_{\mu\nu}X^\mu X^\nu=0$, it can be shown that
\begin{equation}\label{tt5}
J_{\mu}\Gamma^\mu_{\alpha\beta} \frac{dX^\alpha}{d\varrho} \frac{dX^\beta}{d\varrho}= \frac{dX^\alpha}{d\varrho} \frac{dX^\beta}{d\varrho} X^\lambda\frac{\partial g_{\alpha\lambda}}{\partial X^\beta}+ \frac{dX^\alpha}{d\varrho} \frac{dX^\beta}{d\varrho} X^\lambda\frac{\partial f_{\alpha\lambda}}{\partial X^\beta}
\end{equation}
\noindent Summing Eq.~(\ref{tt4}) with Eq.~(\ref{tt5}) and considering Eq.~(\ref{tt3}), we find
\begin{equation}\label{tt6}
\frac{d {\tilde  P}}{d\varrho}=\Bigl(\frac{d\varsigma}{d\varrho}\Bigr)^2
\end{equation}
\noindent Integrating Eq.~(\ref{tt6}) from the initial condition to the steady-state, we find
\begin{equation}\label{tt6a}
{\tilde P}(X_s)-{\tilde P}=\int
\Bigl(\frac{d\varsigma}{d\varrho}\Bigr)^2
d\varrho\ge0
\end{equation}
\noindent where the inequality is only saturated at the steady-state. Recalling Eq.~(\ref{tt2b}), we also have 
\begin{equation}\label{tt6ab}
\frac{d\varrho}{dt}{\tilde P}(X_s)=\Bigl[\exp\Bigl(-\int_t\varphi^{*}(t') dt'\Bigr)\Bigr]\ {\mathcal P}(X_s)=0
\end{equation}
\noindent Hence, the inequality established by the UCE can be derived
\begin{equation}\label{tt8}
{\mathcal P}={\tilde P}\frac{d\varrho}{dt}=J_\mu\frac{dX^\mu}{d\varrho}\Bigl(\exp\int\varphi^{*} dt\Bigr)=-\Bigl(\exp\int\varphi^{*}(t')dt'\Bigr)\int\Bigl(\frac{d\varsigma}{d\varrho}\Bigr)^2d\varrho\le0
\end{equation}
\noindent where Eq.~(\ref{tt6ab}) has been taken into account. Eq.~(\ref{tt6}) can be re-written as 
\begin{equation}\label{tt8a}
\frac{d}{d\varsigma}\Bigl[\Bigl(\frac{d\varsigma}{d\varrho}\Bigr)P\Bigr]=\Bigl(\frac{d\varsigma}{d\varrho}\Bigr)
\end{equation}
\noindent This equation generalizes Eq.~(\ref{i12}), which was valid only in the near-equilibrium region (note that, in the linear region, $d\varsigma/d\varrho=1/b=const.$). Integrating Eq.~(\ref{tt8a}), the expression of the dissipative quantity $P$ is derived
\begin{equation}\label{tt8b}
P-P(\varsigma=l)=-\Bigl(\frac{d\varrho}{d\varsigma}\Bigr)\int_\varsigma^l\Bigl(\frac{d\varsigma'}{d\varrho}\Bigr)d\varsigma'=-{\Bigl(g_{\mu\nu}\frac{dX^\mu}{d\varrho}\frac{dX^\nu}{d\varrho}\Bigr)}^{\!\!-1/2}\!\!\!\!\int_\varsigma^l\!\!{\Bigl(g_{\mu\nu}\frac{dX^\mu}{d\varrho}\frac{dX^\nu}{d\varrho}\Bigr)}^{\!\!1/2}\!\!\!d\varsigma'\le0
\end{equation}
\noindent On the right, it is understood that the $X$'s are expressed in terms of $\varrho(\varsigma)$. Eq.~(\ref{tt8b}) generalizes Eq.~(\ref{i10}), which was valid only in the linear region. Note that, in the Onsager region, the validity of the MEPT requires $P(\varsigma=l)=0$. This because the steady-state corresponds to the state of minimum entropy production. Out of the linear region, this equation may not be satisfied. For $a-a$ or $b-b$ processes in the Onsager region, Eq.~(\ref{tt8}) implies the validity of the inequality (\ref{tt1}). Indeed, Eq.~(\ref{ts4}) gives
\begin{equation}\label{tt9}
\frac{\delta\sigma}{\delta\varrho}=\frac{d\sigma}{d\varrho}=J_\mu\frac{\delta X^\mu}{\delta\varrho}+X^\mu\frac{\delta  J_\mu}{d\varrho}=2J_\mu\frac{\delta X^\mu}{\delta\varrho}+X^\mu X^\nu\frac{\delta L_{\mu\nu}}{\delta\varrho}
\end{equation}
\noindent In the linear region, the coefficients of the affine connection vanish. Eq.~(\ref{tt9}) is simplified reducing to
\begin{equation}\label{tt10}
\frac{d\sigma}{dt}=\frac{d\sigma}{d\varrho}\frac{d\varrho}{dt}=2\Bigl(J_\mu\frac{d X^\mu}{d\varrho}\frac{d\varrho}{dt}\Bigr)=2{\mathcal P}\le0
\end{equation}
\noindent where the inequality is saturated only at the steady state. 

\noindent Let us now consider the relaxation of spatially extended thermodynamic systems. We say that a spatially-extended system {\it relaxes} (from the geometrical point of view) towards a steady-state if the thermodynamic mode (i.e., the mode ${\bf k} = {\bf 0}$) relaxes to the steady-state following the shortest path. In this case, the dissipative quantity should be expressed in the integral form
\begin{equation}\label{tt11}
\mathcal{P}= \int_\Omega {\mathcal J}_\mu({\bf r},t) d_t{\mathcal X}^\mu({\bf r},t)\ dv 
\end{equation}
\noindent where $d_t{\mathcal X}^\mu\equiv d{\mathcal X}^\mu/dt$. In terms of wave-vectors ${\bf k}$, Eq.~(\ref{tt11}) can easily be brought into the form
\begin{equation}\label{tt15}
\mathcal{P}= \Omega \Bigl({\hat J}_{\mu (0)}(t){d_t{\hat X}}^\mu_{(0)}(t)+\sum_{{\bf k}\neq0}
{\hat J}_{\mu ({\bf k})}(t)d_t{\hat  X}^\mu_{({\bf -k})}(t)\Bigr)
\end{equation}
\noindent where Eq.~(\ref{sets6}) has been taken into account. As already mentioned in section \ref{isoef}, in a relaxation process, contributions from different wave-numbers are negligible with respect to those with same wave-numbers \cite{haken}. Hence, recalling Eqs~(\ref{sets4}) and the fact that ${\hat g}_{\mu\nu(0)}(t)$ is a positive definite matrix (see Section \ref{isoef}), we finally obtain 
\begin{equation}\label{tt16}
{\mathcal P}=\int_\Omega {\mathcal J}_\mu({\bf r},t) d_t{\mathcal X}^\mu({\bf r},t)\ dv \simeq \Omega J_{\mu}(t)d_t X^\mu(t)\leq 0
\end{equation}
\noindent
where inequality (\ref{tt8}) has also been taken into account. It is therefore proven that the Universal Criterion of Evolution is automatically satisfied if the system relaxes along the shortest path. Indeed it would be more exact to say: the affine connection, given in Eq.~(\ref{ts9}), has been constructed in such a way that the UCE is satisfied without imposing {\it any} restrictions on the transport coefficients (i.e., on matrices $g_{\mu\nu}$ and $f_{\mu\nu}$). In addition, analogously to Christoffel's symbols, the elements of the new affine connection have been constructed from matrices $g_{\mu\nu}$ and $f_{\mu\nu}$ and their first derivatives in such a way that all coefficients vanish in the Onsager region. Eq.~(\ref{ts9}) provides the simplest expression satisfying these requirements.
\vskip 0.2truecm
\noindent {\bf The Minimum Rate of Dissipation Principle} (MRDP)

In ref.\cite{sonnino1} the validity of the following theorem is shown: 

\noindent {\it The generally covariant part of the Glansdorff-Prigogine quantity is always negative and is locally minimized when the evolution of a system traces out a geodesic in the space of thermodynamic configurations}.

\noindent It is important to stress that this theorem does not refer to the Glansdorff-Prigogine expression reported in Eq.~(\ref{tt2}) but only to its {\it generally covariant part}. Moreover, it concerns the evolution of a system in the space of thermodynamic configurations and {\it not} in the thermodynamic space. One could consider the possibility that the shortest path in the thermodynamic space is an extremal for the functional
\begin{equation}\label{tt17}
\int_{\varsigma_1}^{\varsigma_2}J_\mu{\dot X}^\mu d\varsigma
\end{equation}
\noindent The answer is negative. Indeed, a curve is an extremal for functional Eq.~(\ref{tt17}) if, and only if, it satisfies Euler's equations\footnote{Notice that $J_{\nu,\mu}-J_{\mu,\nu}$ is a thermodynamic tensor of second order. \label{Euler}}
\begin{equation}\label{tt18}
{\dot X}^\nu\Bigl(\frac{\partial J_\nu}{\partial X^\mu}-\frac{\partial J_\mu}{\partial X^\nu}\Bigr)=0
\end{equation}
\noindent 
As it can be easily checked, this extremal coincides with the shortest path if
\begin{eqnarray}\label{tt19}
&&\!\!\!\!\!\!\!\!\!\!\!\!\!\!\!\!  
\frac{1}{2}\Bigl(\frac{M_{\mu\alpha}}{\partial X^\beta}+\frac{M_{\mu\beta}}{\partial X^\alpha}\Bigr)-\Gamma^{\kappa}_{\alpha\beta}M_{\mu\kappa}=0\qquad{\rm where}\\
&&\!\!\!\!\!\!\!\!\!\!\!\!\!\!\!\!
M_{\mu\nu}\equiv J_{\nu,\mu}-J_{\mu,\nu}=2f_{\nu\mu}+X^\kappa(g_{\nu\kappa,\mu}-g_{\mu\kappa,\nu})+X^\kappa(f_{\nu\kappa,\mu}-f_{\mu\kappa,\nu})\nonumber
\end{eqnarray}
\noindent and $\Gamma^{\kappa}_{\alpha\beta}$ given in Eq.~(\ref{ts25}). However, Eqs.~(\ref{tt19}) are $n^2(n+1)/2$ equations for $n^2$ variables (the transport coefficients) and, in general, for $n\neq 1$, they do not admit solutions. We have thus another proof that the Universal Criterion of Evolution can not be derived from a variational principle.

\vskip 0.2truecm
\section{Conclusions and Limit of Validity of the Approach}\label{cs}
\vskip 0.2truecm
\noindent A macroscopic description of thermodynamic systems requires the formulation of a theory for the closure relations. To this purpose, a thermodynamic field theory has been proposed a decade ago. The aim of this theory was to determine the (non linear) deviations from of the Onsager coefficients, which satisfy the thermodynamic theorems for systems out of equilibrium. The Onsager matrix, which depends on the materials under consideration, entered in the theory as an input. Magnetically confined tokamak plasmas are an example of thermodynamic systems where the first basic assumption of the Onsager microscopic theory of fluctuations is not satisfied. This prevents the phenomenological relations from being linear. Another interesting case may be met in hydrodynamics. In some circumstances, indeed, nonlinear terms of convective origin may arise \cite{ottinger}, as for instance in frame-indifferent time derivatives as co-rotational Jaumann derivative or upper-convected Maxwell time derivatives, which do not modify the entropy production. 

The main purpose of this paper is to present a new formulation of the thermodynamical field theory where one of the basic restrictions, namely the closed-form of the skew-symmetric piece of the transport coefficients (see Ref.\cite{sonnino}), has been removed. Furthermore, the general covariance principle, respected, in reality, only by a very limited class of thermodynamic processes, has been replaced by the thermodynamic covariance principle, first introduced by De Donder and Prigogine for treating non equilibrium chemical reactions \cite{dedonder}. The validity of the De Donder-Prigogine statement has been successfully tested, without exception until now, in a wide variety of physical processes going beyond the domain of chemical reactions. The introduction of this principle requested, however, the application of an appropriate mathematical formalism, which may be referred to as the {\it entropy-covariant formalism}. The construction of the present theory rests on two assumptions:
\begin{itemize}
\item The thermodynamic theorems valid when a generic thermodynamic system relaxes out of equilibrium are satisfied;
\item There exists a thermodynamic action, scalar under thermodynamic coordinate transformations, which is stationary for general variations in the transport coefficients and the affine connection.
\end{itemize}
\noindent The second strong assumption can only be judged by its results. A non-Riemannian geometry has been constructed out of the components of the affine connection, which has been determined by imposing the validity of the Universal Criterion of Evolution for non-equilibrium systems relaxing towards a steady-state. Relaxation expresses an intrinsic physical property of a thermodynamic system. The affine connection, on the other hand, is an intrinsic property of geometry allowing to determine the equation for the shortest path. It is the author's opinion that a correct thermodynamical-geometrical theory should correlate these two properties. It is important to mention that the thermodynamic space tends to be Riemannian for small values of the inverse of the entropy production. In this limit, we obtain again the same closure relations found in Ref.\cite{sonnino}. The results established for magnetically confined plasmas \cite{sonnino3}, and for the nonlinear thermoelectric effect and the unimolecular triangular reaction \cite{sonnino4}, remain then valid. 

Finally, note that the transport equations may take even more general forms than Eq.~(\ref{ief0}). The fluxes and the forces can be defined locally as fields depending on space coordinates and time. The most general transport relation takes the form
\begin{equation}\label{c1}
J_{\mu}({\bf r},t)=\int_\Omega d{\bf r'}\int_0^t dt' {\mathcal L}_{\mu\nu}(X({\bf r'},t'))X^\nu({\bf r}-{\bf r'},t-t')
\end{equation}
\noindent This type of nonlocal and non Markovian equation expresses the fact that the flux at a given point (${\bf r},t$) could be influenced by the values of the forces in its spatial environment and by its history. Whenever the spatial and temporal ranges of influence are sufficiently small, the delocalization and the retardation of the forces can be neglected under the integral:
\begin{equation}\label{c1a}
{\mathcal L}_{\mu\nu}(X({\bf r'},t'))X^\nu({\bf r}-{\bf r'},t-t')\simeq 
2\tau_{\mu\nu}(X({\bf r},t))X^\nu({\bf r},t)\delta({\bf r}-{\bf r'})\delta (t-t')
\end{equation}
\noindent where $\delta$ denotes the Dirac delta function. In this case, the transport equations reduces to
\begin{equation}\label{c2}
J_{\mu}({\bf r},t)\simeq\tau_{\mu\nu}(X({\bf r},t))X^\nu({\bf r},t)
\end{equation}
\noindent In the vast majority of cases studied at present in transport theory, it is assumed that the transport equations are of the form of Eq.~(\ref{c2}). However, equations of the form Eq.~(\ref{c1}) may be met when we deal with anomalous transport processes such as, for example, transport in turbulent tokamak plasmas \cite{balescu3}. Eq.~(\ref{c1a}) establishes, in some sort, the limit of validity of the present approach: {\it Eqs~(\ref{tfe1}) determine the nonlinear corrections to the linear ("Onsager") transport coefficients whenever the width of the nonlocal coefficients can be neglected}. It is worthwhile mentioning that in this manuscript, the thermodynamic quantities (number density, temperature, pressure etc.) are evaluated at the local equilibrium state. This is not inconsistent with the fact that the arbitrary state of a thermodynamic system is close to (but not in) a state of local equilibrium. Indeed, as known, it is always possible to construct a representation in such a way that the thermodynamic quantities, evaluated with a distribution function close to a Maxwellian, do coincide {\it exactly} with those evaluated at the local equilibrium state (see, for example, the textbook \cite{balescu1}). 

\vskip 0.2truecm
\section{Acknowledgments}
I would like to pay tribute to the memory of Prof. I. Prigogine who gave me the opportunity to exchange most interesting views in different areas  of thermodynamics of irreversible processes. My strong interest in this domain of research is due to him, who promoted the {\it Brussels School of Thermodynamics} at the U.L.B., where I took my doctorate in Physics. I am also very grateful to Prof. M. Malek Mansour and Prof. M. Tlidi, from the Universit{\'e} Libre de Bruxelles, Prof. C.M. Becchi and Prof. E. Massa, from the University of Genoa, Dr. F. Zonca, from the EURATOM/ENEA Italian Fusion Association in Frascati (Rome) and Dr. J. Evslin from the SISSA (International School for Advanced Studies) for the useful discussions and suggestions. I would like to thank my hierarchy at the European Commission and the members of the Association-Belgian State for Controlled Thermonuclear Fusion at the U.L.B.
\vskip 0.2truecm
\appendix 
\noindent \section{Transformation Law and Properties of the Affine Connection Eq.(\ref{ts25}).}\label{appx}
\vskip 0.2truecm
In this section we show that the affine connection Eq.~(\ref{ts25}) transforms, under TCT, as in Eq.~(\ref{ts1}) and satisfies the postulates ${\bf 1.}$, ${\bf 2.}$ and ${\bf 3.}$ We first note that the quantity $\delta_\alpha^\lambda\psi_{\beta}+\delta_\beta^\lambda\psi_{\alpha}$ transforms like a mixed thermodynamic tensor of third rank
\begin{equation}\label{appx0}
\delta_\alpha^\lambda\psi_{\beta}+\delta_\beta^\lambda\psi_{\alpha}=(
\delta_\rho^\tau\psi_{\nu}+\delta_\nu^\tau\psi_{\rho})
\frac{\partial X'^\lambda}{\partial X^\tau}\frac{\partial X^\rho}{\partial X'^\alpha}\frac{\partial X^\nu}{\partial X'^\beta}
\end{equation}
\noindent Thus, if Eq.~(\ref{ts9}) transforms, under TCT, like Eq.~(\ref{ts1}), then so will be Eq.~(\ref{ts25}). Consider the symmetric processes. From Eq.~(\ref{tr4}), we have
\begin{equation}\label{appx1}
\!\!\frac{\partial g'_{\alpha\beta}}{\partial X'^\kappa}\!=\!\frac{\partial g_{\rho\nu}}{\partial X^\varrho}\frac{\partial X^\varrho}{\partial X'^\kappa}\frac{\partial X^\rho}{\partial X'^\alpha}\frac{\partial X^\nu}{\partial X'^\beta}+g_{\rho\nu}\frac{\partial^2X^\rho}{\partial X'^\kappa\partial X'^\alpha}\frac{\partial X^\nu}{\partial X'^\beta}+g_{\rho\nu}\frac{\partial^2X^\rho}{\partial X'^\kappa\partial X'^\beta}\frac{\partial X^\nu}{\partial X'^\alpha}\\
\end{equation}
\noindent The thermodynamic Christoffel symbols transform then as 
\begin{equation}\label{appx3}
\begin{Bmatrix} 
\lambda \\ \alpha\beta
\end{Bmatrix}'=
\begin{Bmatrix} 
\tau \\ \rho\nu
\end{Bmatrix}\frac{\partial X'^\lambda}{\partial X^\tau}\frac{\partial X^\rho}{\partial X'^\alpha}\frac{\partial X^\nu}{\partial X'^\beta}+\frac{\partial X'^\lambda}{\partial X^\rho}\frac{\partial^2X^\rho}{\partial X'^\alpha\partial X'^\beta}
\end{equation}
\noindent Recalling that $\sigma'=\sigma$, from Eq.~(\ref{appx1}) we also find
\begin{eqnarray}\label{appx4}
&&\frac{{\check N}^{'\lambda\kappa}}{2\sigma'}X'_\kappa\mathcal{O}'(g'_{\alpha\beta})=\frac{{\check N}^{\tau\eta}}{2\sigma}X_\eta\mathcal{O}(g_{\rho\nu})\frac{\partial X'^\lambda}{\partial X^\tau}\frac{\partial X^\rho}{\partial X'^\alpha}\frac{\partial X^\nu}{\partial X'^\beta}\\
&&\frac{1}{2\sigma'}X'^\lambda\mathcal{O}'(g'_{\alpha\beta})=\frac{1}{2\sigma}X^\tau\mathcal{O}(g_{\rho\nu})\frac{\partial X'^\lambda}{\partial X^\tau}\frac{\partial X^\rho}{\partial X'^\alpha}\frac{\partial X^\nu}{\partial X'^\beta}\nonumber
\end{eqnarray}
\noindent where Eqs~(\ref{tr2}) and Eqs~(\ref{tct2}) have been taken into account. Therefore, the affine connection 
\begin{equation}\label{appx5}
{\Gamma}_{\rho\nu}^\tau=
\begin{Bmatrix} 
\tau \\ \rho\nu
\end{Bmatrix}+
\frac{1}{2\sigma}X^\tau\mathcal{O}(g_{\rho\nu})-\frac{1}{2(n+1)\sigma}[\delta_\rho^\tau X^\eta\mathcal{O}(g_{\nu\eta})+\delta_\nu^\tau X^\eta\mathcal{O}(g_{\rho\nu})]
\end{equation}
\noindent transforms as 
\begin{equation}\label{appx6}
{\Gamma'}_{\alpha\beta}^\lambda=
\Gamma_{\rho\nu}^\tau
\frac{\partial X'^\lambda}{\partial X^\tau}\frac{\partial X^\rho}{\partial X'^\alpha}\frac{\partial X^\nu}{\partial X'^\beta}+\frac{\partial X'^\lambda}{\partial X^\rho}\frac{\partial^2X^\rho}{\partial X'^\alpha\partial X'^\beta}
\end{equation}
Consider now the general case. From Eq.~(\ref{tr4}) we obtain
\begin{eqnarray}\label{appx7}
\frac{1}{2}\Bigl(\frac{\partial g'_{\alpha\kappa}}{\partial X'^\beta}+\frac{\partial g'_{\beta\kappa}}{\partial X'^\beta}-\frac{\partial g'_{\alpha\beta}}{\partial X'^\kappa}\Bigr)=\!\!\!\!&&\frac{\partial X^\varrho}{\partial X'^\kappa}\frac{\partial X^\rho}{\partial X'^\alpha}\frac{\partial X^\nu}{\partial X'^\beta}
\Bigl[\frac{1}{2}\Bigl(\frac{\partial g_{\nu\varrho}}{\partial X^\rho}+\frac{\partial g_{\rho\varrho}}{\partial X^\nu}-\frac{\partial g_{\rho\nu}}{\partial X^\varrho}\Bigr)\Bigr] \nonumber\\
\!\!\!\!\!\!\!\!&&+g_{\rho\nu}\frac{\partial^2 X^\rho}{\partial X'^\alpha\partial X'^\beta}\frac{\partial X^\nu}{\partial X'^\kappa}
\end{eqnarray}
\noindent From Eq.~(\ref{tr6}), we also have
\begin{eqnarray}\label{appx8}
&& \!\!\!\!\frac{\partial f'_{\alpha\mu}}{\partial X'^\beta}\!=\!\frac{\partial f_{\rho\eta}}{\partial X^\varsigma}\frac{\partial X^\varsigma}{\partial X'^\beta}\frac{\partial X^\rho}{\partial X'^\alpha}\frac{\partial X^\eta}{\partial X'^\mu}\!+\! f_{\rho\eta}\frac{\partial^2X^\rho}{\partial X'^\beta\partial X'^\alpha}\frac{\partial X^\eta}{\partial X'^\mu}\!+\!f_{\rho\eta}\frac{\partial^2X^\rho}{\partial X'^\beta\partial X'^\mu}\frac{\partial X^\eta}{\partial X'^\alpha}
\nonumber\\
&&\!\!\!\!\frac{\partial f'_{\beta\mu}}{\partial X'^\alpha}\!=\!\frac{\partial f_{\varsigma\eta}}{\partial X^\rho}\frac{\partial X^\varsigma}{\partial X'^\beta}\frac{\partial X^\rho}{\partial X'^\alpha}\frac{\partial X^\eta}{\partial X'^\mu}\!+\! f_{\rho\eta}\frac{\partial^2X^\rho}{\partial X'^\alpha\partial X'^\beta}\frac{\partial X^\eta}{\partial X'^\mu}\!+\!f_{\rho\eta}\frac{\partial^2X^\rho}{\partial X'^\alpha\partial X'^\mu}\frac{\partial X^\eta}{\partial X'^\beta}\nonumber\\
\end{eqnarray}
\noindent Taking into account Eqs~(\ref{tr2}) and Eqs~(\ref{tct2}) we find
\begin{eqnarray}\label{appx9}
\!\!\!\!\!\!\!\!\!\!\!\!\!\!\!\!&& X'_\kappa X'^\mu\frac{\partial f'_{\alpha\mu}}{\partial X'^\beta}=X_\varrho X^\eta\frac{\partial f_{\rho\eta}}{\partial X^\nu}\frac{\partial X^\varrho}{\partial X'^\kappa}\frac{\partial X^\rho}{\partial X'^\alpha}\frac{\partial X^\nu}{\partial X'^\beta}+X_\nu X^\eta f_{\rho\eta}\frac{\partial^2 X^\rho}{\partial X'^\alpha\partial X'^\beta}\frac{\partial X^\nu}{\partial X'^\kappa} \nonumber\\
\!\!\!\!\!\!\!\!\!\!\!\!\!\!\!\!&&X'_\kappa X'^\mu\frac{\partial f'_{\beta\mu}}{\partial X'^\alpha}=X_\varrho X^\eta\frac{\partial f_{\nu\eta}}{\partial X^\rho}\frac{\partial X^\varrho}{\partial X'^\kappa}\frac{\partial X^\rho}{\partial X'^\alpha}\frac{\partial X^\nu}{\partial X'^\beta}+X_\nu X^\eta f_{\rho\eta}\frac{\partial^2 X^\rho}{\partial X'^\alpha\partial X'^\beta}\frac{\partial X^\nu}{\partial X'^\kappa}\nonumber\\
\end{eqnarray}
\noindent from which we obtain
\begin{eqnarray}\label{appx10}
\frac{1}{2\sigma'}X'_\kappa X'^\mu\Bigl(\frac{\partial f'_{\alpha\mu}}{\partial X'^\beta}+\frac{\partial f'_{\beta\mu}}{\partial X'^\alpha}\Bigr)=\!\!\!&&\frac{\partial X^\varrho}{\partial X'^\kappa}\frac{\partial X^\rho}{\partial X'^\alpha}\frac{\partial X^\nu}{\partial X'^\beta}\Bigl[\frac{1}{2\sigma}X_\varrho X^\eta\Bigl(\frac{\partial f_{\rho\eta}}{\partial X^\nu}+\frac{\partial f_{\nu\eta}}{\partial X^\rho}\Bigr)\Bigr]\nonumber\\
&& +\frac{1}{\sigma}X_\nu X^\eta f_{\rho\eta}\frac{\partial^2 X^\rho}{\partial X'^\alpha\partial X'^\beta}\frac{\partial X^\nu}{\partial X'^\kappa}
\end{eqnarray}
\noindent Let us now re-consider the transformations of the following quantities
\begin{eqnarray}\label{appx11}
&& \!\!\!\!\frac{\partial g'_{\alpha\mu}}{\partial X'^\beta}\!=\!\frac{\partial g_{\rho\eta}}{\partial X^\varsigma}\frac{\partial X^\varsigma}{\partial X'^\beta}\frac{\partial X^\rho}{\partial X'^\alpha}\frac{\partial X^\eta}{\partial X'^\mu}\!+\! g_{\rho\eta}\frac{\partial^2X^\rho}{\partial X'^\beta\partial X'^\alpha}\frac{\partial X^\eta}{\partial X'^\mu}\!+\!g_{\rho\eta}\frac{\partial^2X^\rho}{\partial X'^\beta\partial X'^\mu}\frac{\partial X^\eta}{\partial X'^\alpha}
\nonumber\\
&&\!\!\!\!\frac{\partial g'_{\beta\mu}}{\partial X'^\alpha}\!=\!\frac{\partial g_{\varsigma\eta}}{\partial X^\rho}\frac{\partial X^\varsigma}{\partial X'^\beta}\frac{\partial X^\rho}{\partial X'^\alpha}\frac{\partial X^\eta}{\partial X'^\mu}\!+\! g_{\rho\eta}\frac{\partial^2X^\rho}{\partial X'^\alpha\partial X'^\beta}\frac{\partial X^\eta}{\partial X'^\mu}\!+\!g_{\rho\eta}\frac{\partial^2X^\rho}{\partial X'^\alpha\partial X'^\mu}\frac{\partial X^\eta}{\partial X'^\beta}\nonumber\\
\end{eqnarray}
\noindent From these equations we obtain
\begin{equation}\label{appx12}
X'^\mu\frac{\partial g'_{\alpha\mu}}{\partial X'^\beta}+X'^\mu\frac{\partial g'_{\beta\mu}}{\partial X'^\alpha}\!=\!\Bigl(X^\eta\frac{\partial g_{\rho\eta}}{\partial X^\varsigma}+
\frac{\partial g_{\varsigma\eta}}{\partial X^\rho}\Bigr)\frac{\partial X^\varsigma}{\partial X'^\beta}\frac{\partial X^\rho}{\partial X'^\alpha}\frac{\partial X^\eta}{\partial X'^\mu}\!+\! 2X_\rho\frac{\partial^2X^\rho}{\partial X'^\beta\partial X'^\alpha}
\end{equation}
\noindent where Eqs~(\ref{tct2}) have been taken into account. From Eq.~(\ref{appx12}) we finally obtain
\begin{eqnarray}\label{appx13}
&&\!\!\!\!\!\!\!\!\!\!\!\!\!\!\!\!\!\!\!\!\!\!\!\!\!\!\!\!\frac{1}{2\sigma'}\Bigl[X'^\varsigma\Bigl(\frac{\partial g'_{\alpha\varsigma}}{\partial X'^\beta}+\frac{\partial g'_{\beta\varsigma}}{\partial X'^\alpha}\Bigr)\Bigr]f'_{\kappa\mu}X'^\mu=\nonumber\\
&&\qquad\qquad\quad\frac{1}{2\sigma}\Bigl[X^\eta\Bigl(\frac{\partial g_{\rho\eta}}{\partial X^\nu}+\frac{\partial g_{\nu\eta}}{\partial X^\rho}\Bigr)\Bigr]f_{\varrho\varsigma}X^\varsigma\frac{\partial X^\varrho}{\partial X'^\kappa}\frac{\partial X^\rho}{\partial X'^\alpha}\frac{\partial X^\nu}{\partial X'^\beta}
\nonumber\\
&&\qquad\qquad\quad+\frac{1}{\sigma}X_\rho X^\eta f_{\nu\eta}\frac{\partial^2 X^\rho}{\partial X'^\alpha\partial X'^\beta}\frac{\partial X^\nu}{\partial X'^\kappa}
\end{eqnarray}
\noindent Summing Eq.~(\ref{appx7}) with Eqs~(\ref{appx10}) and (\ref{appx13}), it follows that
\begin{equation}\label{appx14}
{\check\Gamma}_{\alpha\beta}^{\prime\lambda}=
{\check\Gamma}_{\rho\nu}^\tau
\frac{\partial X'^\lambda}{\partial X^\tau}\frac{\partial X^\rho}{\partial X'^\alpha}\frac{\partial X^\nu}{\partial X'^\beta}+\frac{\partial X'^\lambda}{\partial X^\rho}\frac{\partial^2 X^\rho}{\partial X'^\alpha\partial X'^\beta}
\end{equation}
\noindent where 
\begin{eqnarray}\label{appx15}
\!\!\!\!\!\!\!
{\check\Gamma}_{\rho\nu}^\tau=\!\!\!\!\!&&{\check N}^{\tau\varrho}g_{\varrho\varsigma}\begin{Bmatrix} 
\varsigma \\ \rho\nu
\end{Bmatrix}+\frac{{\check N}^{\tau\varrho} X_{\varrho}X^\eta}{2\sigma}\Bigl(\frac{\partial f_{\rho\eta}}{\partial X^\nu}+\frac{\partial f_{\nu\eta}}{\partial X^\rho}\Bigr)
\nonumber\\
&&\qquad\qquad\qquad\qquad\qquad\qquad
+\frac{{\check N}^{\tau\varrho}f_{\varrho\varsigma}X^\varsigma X^\eta}{2\sigma}
\Bigr(\frac{\partial g_{\rho\eta}}{\partial X^\nu}+\frac{\partial g_{\nu\eta}}{\partial X^\rho}\Bigl)
\end{eqnarray}
\noindent and 
\begin{equation}\label{appx16}
{\check N}^{\tau\varrho}N_{\rho\varrho}=\delta_{\rho}^{\tau}\qquad {\rm}\quad {\rm with}\quad N_{\rho\varrho}=g_{\rho\varrho}+\frac{1}{\sigma}f_{\rho\eta}X^\eta X_\varrho+\frac{1}{\sigma}f_{\varrho\eta}X^\eta X_\rho
\end{equation}
\noindent Summing again Eq.~(\ref{appx15}) with Eq.~(\ref{appx0}) and the first equation of Eq.~(\ref{appx4}), we finally obtain
\begin{equation}\label{appx17}
{\Gamma'}_{\alpha\beta}^\lambda=
\Gamma_{\rho\nu}^\tau
\frac{\partial X'^\lambda}{\partial X^\tau}\frac{\partial X^\rho}{\partial X'^\alpha}\frac{\partial X^\nu}{\partial X'^\beta}+\frac{\partial X'^\lambda}{\partial X^\rho}\frac{\partial^2X^\rho}{\partial X'^\alpha\partial X'^\beta}
\end{equation}
\noindent where
\begin{equation}\label{appx18}
\Gamma_{\rho\nu}^\tau={\check\Gamma}_{\rho\nu}^\tau+\frac{{\check N}^{\tau\eta}}{2\sigma}X_\eta\mathcal{O}(g_{\rho\nu})+\delta_\rho^\tau\psi_{\nu}+\delta_\nu^\tau\psi_{\rho}
\end{equation}
\noindent It is not difficult to prove that the affine connection Eq.~(\ref{ts25}) satisfies the postulates ${\bf 1.}$, ${\bf 2.}$ and ${\bf 3.}$ Indeed, if $A^\mu$ indicates a thermodynamic vector, we have 
\begin{equation}\label{appx19}
A'^\lambda=A^\eta\frac{\partial X'^\lambda}{\partial X^\eta}
\end{equation}
\noindent Deriving this equation, with respect to parameter $\varsigma$, we obtain
\begin{equation}\label{appx20}
\frac{dA'^\lambda}{d\varsigma}=\frac{dA^\eta}{d\varsigma}\frac{\partial X'^\lambda}{\partial X^\eta}+A^\eta\frac{\partial^2 X'^\lambda}{\partial X^\tau\partial X^\eta}\frac{dX^\tau}{d\varsigma}
\end{equation}
\noindent Taking into account the following identities 
\begin{equation}\label{appx20a}
\frac{\partial^2 X'^\lambda}{\partial X^\tau\partial X^\eta}=-\frac{\partial X'^\lambda}{\partial X^\rho}\frac{\partial X'^\alpha}{\partial X^\tau}\frac{\partial^2 X^\rho}{\partial X^\eta\partial X'^\alpha}=-\frac{\partial X'^\alpha}{\partial X^\tau}\frac{\partial X'^\beta}{\partial X^\eta}\frac{\partial X'^\lambda}{\partial X^\rho}\frac{\partial^2 X^\rho}{\partial X'^\alpha\partial X'^\beta}
\end{equation}
\noindent and Eq.~(\ref{appx17}), we find
\begin{equation}\label{appx21}
\frac{\delta A'^\lambda}{\delta \varsigma}=\frac{\delta A^\eta}{\delta \varsigma}\frac{\partial X'^\lambda}{\partial X^\eta}
\end{equation}
\noindent The validity of postulates ${\bf 2.}$ and ${\bf 3.}$ is immediately verified, by direct computation, using Eqs~(\ref{ts2}) and (\ref{ts5}). The validity of these postulates was shown above for a thermodynamic vector. By a closely analogous procedure it can be checked that the postulated ${\bf 1.}$ , ${\bf 2.}$ and ${\bf 3.}$ are satisfied for any thermodynamic tensor.
\vskip 0.2truecm
\noindent \section{Derivation of the Nonlinear Closure Equations from the Action Principle.}\label{appeq}
\vskip 0.2truecm
In this appendix, the nonlinear closure equations by the principle of the least action are derived. Let us rewrite Eq.~(\ref{pa5}) as
\begin{equation}\label{appeq1}
I=\int\Bigl[ R_{\mu\nu}g^{\mu\nu}-(\Gamma^\lambda_{\mu\nu}-
{\tilde\Gamma}^\lambda_{\mu\nu})S^{\mu\nu}_\lambda
\Bigr]\sqrt{g}\ \! {d^{}}^n\!X
\end{equation}
\noindent where the expression of $S^{\mu\nu}_\lambda$ is given by Eq.~(\ref{pa4}). This action is stationary by varying independently the transport coefficients (i.e. by varying, separately, $g_{\mu\nu}$ and $f_{\mu\nu}$) and the affine connection $\Gamma^\lambda_{\mu\nu}$. A variation with respect to $\Gamma^\lambda_{\mu\nu}$ reads
\begin{equation}\label{appeq2}
\delta I_{\Gamma}=\int\Bigl[ \delta R_{\mu\nu}g^{\mu\nu}-\delta\Gamma^\lambda_{\mu\nu}S^{\mu\nu}_\lambda
\Bigr]\sqrt{g}\ \! {d^{}}^n\!X=0
\end{equation}
\noindent By direct computation, we can check that
\begin{equation}\label{appeq3}
\delta R_{\mu\nu}=(\delta\Gamma^\lambda_{\mu\lambda})_{\mid\nu}-(\delta\Gamma^\lambda_{\mu\nu})_{\mid\lambda}
\end{equation}
\noindent Defining $\mathcal{K}^{\mu\nu}\equiv\sqrt{g}g^{\mu\nu}$, we have the identities
\begin{eqnarray}\label{appeq4}
&&(\mathcal{K}^{\mu\nu}\delta\Gamma^\lambda_{\mu\lambda})_{\mid\nu}=\mathcal{K}^{\mu\nu}_{\mid\nu}\delta\Gamma^\lambda_{\mu\lambda}+\mathcal{K}^{\mu\nu}\delta\Gamma^\lambda_{\mu\lambda\mid\nu}\nonumber\\
&&
(\mathcal{K}^{\mu\nu}\delta\Gamma^\lambda_{\mu\nu})_{\mid\lambda}=\mathcal{K}^{\mu\nu}_{\mid\lambda}\delta\Gamma^\lambda_{\mu\nu}+\mathcal{K}^{\mu\nu}\delta\Gamma¬\lambda_{\mu\nu\mid\lambda}
\end{eqnarray}
\noindent Eq.~(\ref{appeq2}) can be rewritten as
\begin{eqnarray}\label{appeq5}
\delta I_{\Gamma}=&&\!\!\!\!\!
\int(\mathcal{K}^{\mu\nu}\delta\Gamma^\lambda_{\mu\lambda})_{\mid\nu}{d^{}}^n\!X-\int\mathcal{K}^{\mu\nu}_{\mid\nu}\delta\Gamma^\lambda_{\mu\lambda}{d^{}}^n\!X+\int
\mathcal{K}^{\mu\nu}_{\mid\lambda}\delta\Gamma^\lambda_{\mu\nu}{d^{}}^n\!X-
\nonumber\\
&&\!\!\!\!\!\!\!\!\!\!\!\!
\int(\mathcal{K}^{\mu\nu}\delta\Gamma^\lambda_{\mu\nu})_{\mid\lambda}{d^{}}^n\!X
-\int S^{\mu\nu}_\lambda\delta\Gamma^\lambda_{\mu\nu}
\sqrt{g}\ \! {d^{}}^n\!X=0
\end{eqnarray}
\noindent The thermodynamic covariant derivative of the metric tensor reads
\begin{equation}\label{appeq14}
g_{\alpha\beta\mid\lambda}=g_{\alpha\beta,\lambda}-\Gamma^\eta_{\alpha\lambda}g_{\eta\beta}-\Gamma^\eta_{\beta\lambda}g_{\eta\alpha}
\end{equation}
\noindent from which we find
\begin{equation}\label{appeq15}
\Gamma^\beta_{\lambda\beta}=-\frac{1}{2}g^{\alpha\beta}g_{\alpha\beta\mid\lambda}+\frac{1}{2}g^{\alpha\beta}g_{\alpha\beta,\lambda}
\end{equation}
\noindent Taking into account that $\delta\sqrt{g}=1/2\sqrt{g}g^{\mu\nu}\delta g_{\mu\nu}$, Eq.~(\ref{appeq15}) can also be brought into the form 
\begin{equation}\label{appeq15a}
\Gamma^\beta_{\lambda\beta}-\frac{1}{\sqrt{g}}\sqrt{g}_{,\lambda}+\frac{1}{\sqrt{g}}\sqrt{g}_{\mid\lambda}=0
\end{equation}
\noindent On the other hand, we can easily check the validity of the following identities
\begin{eqnarray}\label{appeq7}
&&\!\!\!\!\!\!\!\!\!\!\!\!\!\!\!\!
(\mathcal{K}^{\mu\nu}\delta\Gamma^\lambda_{\mu\lambda})_{\mid\nu}=(\mathcal{K}^{\mu\nu}\delta\Gamma^\lambda_{\mu\lambda})_{,\nu}
+(\Gamma^\beta_{\nu\beta}-\frac{1}{\sqrt{g}}\sqrt{g}_{,\nu}+\frac{1}{\sqrt{g}}\sqrt{g}_{\mid\nu}){\mathcal{K}}^{\mu\nu}\delta\Gamma^\lambda_{\mu\lambda}
\nonumber\\
&&\!\!\!\!\!\!\!\!\!\!\!\!\!\!\!\!
(\mathcal{K}^{\mu\nu}\delta\Gamma^\lambda_{\mu\nu})_{\mid\lambda}=(\mathcal{K}^{\mu\nu}\delta\Gamma^\lambda_{\mu\nu})_{,\lambda}
+(\Gamma^\beta_{\lambda\beta}-\frac{1}{\sqrt{g}}\sqrt{g}_{,\lambda}+\frac{1}{\sqrt{g}}\sqrt{g}_{\mid\lambda}){\mathcal{K}}^{\mu\nu}\delta\Gamma^\lambda_{\mu\nu}
\end{eqnarray}
\noindent Therefore, from Eq.~(\ref{appeq15a}), the terms
\begin{equation}\label{appeq6}
\int(\mathcal{K}^{\mu\nu}\delta\Gamma^\lambda_{\mu\lambda})_{\mid\nu}{d^{}}^n\!X
\qquad{\rm and}\qquad \int(\mathcal{K}^{\mu\nu}\delta\Gamma^\lambda_{\mu\nu})_{\mid\lambda}{d^{}}^n\!X
\end{equation}
\noindent drop out when we integrate over all thermodynamic space. Eq.~(\ref{appeq5}) reduces then to
\begin{equation}\label{appeq9}
\delta I_{\Gamma}=
-\int\mathcal{K}^{\mu\nu}_{\mid\nu}\delta\Gamma^\lambda_{\mu\lambda}{d^{}}^n\!X+\int
\mathcal{K}^{\mu\nu}_{\mid\lambda}\delta\Gamma^\lambda_{\mu\nu}{d^{}}^n\!X-\int S^{\mu\nu}_\lambda\delta\Gamma^\lambda_{\mu\nu}
\sqrt{g}\ \! {d^{}}^n\!X=0
\end{equation}
\noindent It is seen that $\delta I_\Gamma$ vanishes for general variation of $\delta\Gamma^\lambda_{\mu\nu}$ if, and only if, 
\begin{equation}\label{appeq10}
-\frac{1}{2}\mathcal{K}^{\mu\alpha}_{\mid\alpha}\delta^\nu_\lambda
-\frac{1}{2}\mathcal{K}^{\nu\alpha}_{\mid\alpha}\delta^\mu_\lambda+
\mathcal{K}^{\mu\nu}_{\mid\lambda}-
S^{\mu\nu}_\lambda\sqrt{g}=0
\end{equation}
\noindent Contracting indexes $\nu$ with $\lambda$, we find
\begin{equation}\label{appeq11}
\mathcal{K}^{\mu\alpha}_{\mid\alpha}-\Psi^\mu_{\alpha\beta}g^{\alpha\beta}\sqrt{g}=0
\end{equation}
\noindent where Eq.~(\ref{pa4}) has been taken into account. Thanks to Eq.~(\ref{appeq11}), Eq.~(\ref{appeq10}) becomes
\begin{equation}\label{appeq12}
\mathcal{K}^{\mu\nu}_{\mid\lambda}=\Psi^\mu_{\alpha\lambda}g^{\nu\alpha}\sqrt{g}+\Psi^\nu_{\alpha\lambda}g^{\mu\alpha}\sqrt{g}
\end{equation}
\noindent From the identity $\delta g^{\mu\nu}=-g^{\mu\alpha}g^{\nu\beta}\delta g_{\alpha\beta}$, we also have 
\begin{equation}\label{appeq13}
\mathcal{K}^{\mu\nu}_{\mid\lambda}=\sqrt{g}_{\mid\lambda}g^{\mu\nu}+\sqrt{g}g^{\mu\nu}_{\mid\lambda}=\frac{1}{2}\sqrt{g}g^{\mu\nu}g^{\alpha\beta}g_{\alpha\beta\mid\lambda}-\sqrt{g}g^{\mu\alpha}g^{\nu\beta}g_{\alpha\beta\mid\lambda}
\end{equation}
\noindent Eq.~(\ref{appeq12}) reads then
\begin{equation}\label{appeq16}
-g^{\mu\alpha}g^{\nu\beta}g_{\alpha\beta\mid\lambda}+\frac{1}{2}g^{\alpha\beta}g_{\alpha\beta\mid\lambda}g^{\mu\nu}=
\Psi^\mu_{\alpha\lambda}g^{\nu\alpha}+\Psi^\nu_{\alpha\lambda}g^{\mu\alpha}
\end{equation}
\noindent Contracting this equation with $g_{\mu\nu}$, we find, for $n\neq2$
\begin{equation}\label{appeq17}
g^{\alpha\beta}g_{\alpha\beta\mid\lambda}=0
\end{equation}
\noindent where Eqs~(\ref{pa2}) have been taken into account. Eq.~(\ref{appeq16}) is simplified as
\begin{equation}\label{appeq18}
-g^{\mu\alpha}g^{\nu\beta}g_{\alpha\beta\mid\lambda}=
\Psi^\mu_{\alpha\lambda}g^{\nu\alpha}+\Psi^\nu_{\alpha\lambda}g^{\mu\alpha}
\end{equation}
\noindent Contracting again Eq.~(\ref{appeq18}) with $g_{\mu\eta}g_{\nu\rho}$, we finally obtain
\begin{equation}\label{appeq19}
g_{\eta\rho\mid\lambda}=-\Psi^\alpha_{\eta\lambda}g_{\alpha\rho}-\Psi^\alpha_{\rho\lambda}g_{\alpha\eta}
\end{equation}
\noindent The first two equations in Eqs~(\ref{tfe1}) are straightforwardly obtained considering that from Eq.~(\ref{appeq19}) we derive $\Gamma^\lambda_{\mu\nu}-{\tilde \Gamma}^\lambda_{\mu\nu}=0$ (see section \ref{isoef}).
\vskip 0.2truecm
\noindent \section{Comparison between the General Relativity and the Thermodynamic Field Theory Geometries.}\label{comparison}
\vskip 0.2truecm
Although the mathematical symbols are similar, the geometries of the General Relativity and of the TFT are quite different. Above all, in the former case, the geometry is pseudo-Riemannian whereas in the latter is Non-Riemannian. The principle of {\it General Covariance}, respected in the General relativity, is not satisfied in the TFT. In addition, the {\it Equivalence Principle} is not respected in the TFT. On the contrary, the {\it Universal Criterion of Evolution} is satisfied only in the TFT. In the TFT, symbol $R_{\nu\lambda\kappa}^\mu$ should not be confused with the Riemannian curvature tensor and the curvature scalar is defined to as the contraction between the $R_{\nu\lambda}$ thermodynamic tensor (which does not coincide with Ricci's tensor) and the symmetric piece of the transport coefficients (see also Ref. \cite{eisenhart}). In the manuscript it is mentioned that in case of (but only in this case) the dimensionless entropy production is much greater than unity, then the space tends to be Riemannian. However, also in this limit case, a comparison with the General Relativity geometry is not appropriate. The table reported below, should help to avoid any possibility of confusion.
\vskip0.6truecm
\noindent \begin{tabular}{|l|l|l|}
\hline
 & {\bf \ \ General Relativity} &\qquad\qquad\quad  {\bf TFT}\\ \hline
{\bf Geometry} & Pseudo-Riemannian & Non Riemannian\\ \hline
{\bf Field} & Symmetric & Asymmetric\\ \hline
{\bf Metric} & Minkowski (3+1) signat. & Positive-definite\\ \hline
{\bf Space} & Pseudo-Riemannian & Thermodynamic space\\ \hline
{\bf Covariance} & General Covar. Princ. & Homog. funct. of first degree\\ \hline
{\bf Equivalence Principle} & Satisfied & Not statisfied\\ \hline
{\bf Univ. Criterion of Evolution} & Not satisfied & Satisfied\\ \hline
{\bf Main Invariant} & Proper time & Entropy production\\ \hline
${\bf \Gamma}^\mu_{\alpha\beta}$ & Levi-Civita's connection &New thermod. affine connection\\ \hline
${\bf R}^\mu_{\nu\lambda\kappa}$ & Riemannian's tensor & New thermod. curvat. tensor\\ \hline
${\bf R}_{\nu\lambda}$ & Ricci's tensor & New thermod. tensor\\ \hline
${\bf R_{\mu\nu}-1/2g}_{\mu\nu}{\bf R}$ & Einstein's tensor & New thermod. tensor\\ 
\hline
\end{tabular}
\vskip 0.2truecm
\noindent \section{Descriptions of the Mathematical Terms}\label{dictionary}
\vskip 0.2truecm
\noindent For easy reference, we provide below a table with short descriptions of the terms appearing in the manuscript. This should help to make more readable the paper and we refer the reader to the specialized textbooks for rigorous definitions.
\noindent \begin{tabular}{|l|l|l|}
\hline
\qquad\qquad\qquad\ \ {\bf Term} & {\bf \qquad\qquad\quad Description} \\ \hline
{\bf Thermod. Coord. Transf. (TCT)} & $X'^\mu=X^1F^\mu\Bigl(\frac{X^2}{X^1},\ \frac{X^3}{X^2},\ \cdots\ \frac{X^n}{X^{n-1}}\Bigr)$   \\ & where $F^\mu$ are arbitrary functions.\\ \hline
{\bf Covariant thermod. vector $A^\mu$} &  A set of quantities transforming, under TCT, as \\ & $A'^\mu=\frac{\partial X'^{\mu}}{\partial X^\nu} A^\nu$ \\ \hline
{\bf Contra-variant thermod. vector $A_\mu$} & A set of quantities transforming, under TCT, as \\ & $A'_\mu=\frac{\partial X^{\nu}}{\partial X'^\mu}A_\nu$  \\ \hline
{\bf Parallel transport} & Moving a vector along a curve without changing\\
& its direction.\\ \hline
{\bf Affine connection} &  A rule for parallel transport. \\ \hline
{\bf Manifold} & A set of points, which has a continuous $1-1$ map\\ 
&{\it onto} a set of $R^n$.\\ \hline
{\bf Differential manifold} &  A manifold with some additional structure allowing \\ & to do differential calculus on the manifold. \\ \hline
{\bf Linear connection} &  A differential-geometric structure on a differential \\& manifold $\mathcal M$ associated with an affine connection \\ &  on $\mathcal M$, which satisfies the transformation law Eq.~(\ref{ts1}). \\ \hline
{\bf Thermod. affine connection $\Gamma_{\lambda\kappa}^\nu$} &  The affine connection given in Eq.~(\ref{ts25}). \\ \hline
{\bf Tangent space} & A real vector space, containing all possible directions, \\ & attached to every point of a differential manifold. \\ \hline
{\bf Riemannian geometry} & A geometry constructed out of a symmetric, positive \\& definite, second rank tensor.    \\ \hline
{\bf Riemannian manifold} & A real differential manifold in which each tangent \\ & space is equipped with an inner product, which
\\& varies smoothly from point to point. The metric is a \\& positive definite metric tensor.\\ \hline
{\bf Riemannian\ space} & A space equipped with a positive definite metric \\ & tensor and with the Levi-Civita connection. \\ \hline
${\bf Non-Riemannian\ geometry}$ & A geometry constructed out of the components of the \\ & affine connections. \\ \hline
${\bf Thermodynamic\ space}$ & A space equipped with $g_{\mu\nu}$ as metric tensor and with \\ & the  single affine connection given in Eq.~(\ref{ts25}).  \\ 
\hline
\end{tabular}




\end{document}